\documentclass[aps,prb,twocolumn,superscriptaddress]{revtex4-2}
\usepackage{graphicx}
\usepackage{color,amssymb,amsmath}
\usepackage[normalem]{ulem}
\usepackage{hyperref}
\usepackage{bm}

\newcommand{\Deltait}{\mathit{\Delta}}
\newcommand{\Ea}{E_{\mathrm{A}}}
\newcommand{\Phio}{\mathit{\Phi}_0}

\newcommand{\phiL}{\phi_{\mathrm{L}}}
\newcommand{\phiM}{\phi_{\mathrm{M}}}
\newcommand{\phiR}{\phi_{\mathrm{R}}}
\newcommand{\PhiL}{\mathit{\Phi}_{\mathrm{L}}}
\newcommand{\PhiR}{\mathit{\Phi}_{\mathrm{R}}}
\newcommand{\Il}{I_{\mathrm{L}}}
\newcommand{\Ir}{I_{\mathrm{R}}}
\newcommand{\BSO}{\mathbf{B}_\mathrm{SO}}

\newcommand{\Vbias}{V_{\mathrm{bias}}}
\newcommand{\Vtl}{V_{\mathrm{TL}}}
\newcommand{\Vtr}{V_{\mathrm{TR}}}
\newcommand{\Vj}{V_{\mathrm{switch}}}
\newcommand{\Vt}{V_{\mathrm{T}}}
\newcommand{\Vp}{V_{\mathrm{probe}}}
\newcommand{\Vl}{V_{\mathrm{L}}}
\newcommand{\Vm}{V_{\mathrm{M}}}
\newcommand{\Vr}{V_{\mathrm{R}}}

\begin{document}
\title{Spin-degeneracy breaking and parity transitions in three-terminal Josephson junctions}

\author{M.\ Coraiola}
\affiliation{IBM Research Europe---Zurich, 8803 R\"uschlikon, Switzerland}

\author{D.\ Z.\ Haxell}
\affiliation{IBM Research Europe---Zurich, 8803 R\"uschlikon, Switzerland}

\author{D.\ Sabonis}
\affiliation{IBM Research Europe---Zurich, 8803 R\"uschlikon, Switzerland}

\author{M.\ Hinderling}
\affiliation{IBM Research Europe---Zurich, 8803 R\"uschlikon, Switzerland}

\author{S.\ C.\ ten Kate}
\affiliation{IBM Research Europe---Zurich, 8803 R\"uschlikon, Switzerland}

\author{E.\ Cheah}
\affiliation{Laboratory for Solid State Physics, ETH Z\"urich, 8093 Z\"urich, Switzerland}

\author{F.\ Krizek}
\altaffiliation[Present address: ]{Institute of Physics, Czech Academy of Sciences, 162 00 Prague, Czech Republic}
\affiliation{IBM Research Europe---Zurich, 8803 R\"uschlikon, Switzerland}
\affiliation{Laboratory for Solid State Physics, ETH Z\"urich, 8093 Z\"urich, Switzerland}

\author{R.\ Schott}
\affiliation{Laboratory for Solid State Physics, ETH Z\"urich, 8093 Z\"urich, Switzerland}

\author{W.\ Wegscheider}
\affiliation{Laboratory for Solid State Physics, ETH Z\"urich, 8093 Z\"urich, Switzerland}

\author{F.\ Nichele}
\email{fni@zurich.ibm.com}
\affiliation{IBM Research Europe---Zurich, 8803 R\"uschlikon, Switzerland}

\date{\today}

\begin{abstract}
	Harnessing spin and parity degrees of freedom is of fundamental importance for the realization of emergent quantum devices. Nanostructures embedded in superconductor--semiconductor hybrid materials offer novel and yet unexplored routes for addressing and manipulating fermionic modes. Here we spectroscopically probe the two-dimensional band structure of Andreev bound states in a phase-controlled hybrid three-terminal Josephson junction. Andreev bands reveal spin-degeneracy breaking, with level splitting in excess of $\sim {9~\mathrm{GHz}}$, and zero-energy crossings associated to ground state fermion parity transitions, in agreement with theoretical predictions. Both effects occur without the need of external magnetic fields or sizable charging energies and are tuned locally by controlling superconducting phase differences. Our results highlight the potential of multiterminal hybrid devices for engineering quantum states.
\end{abstract}

\maketitle


The spin of quantum particles offers an ideal basis for two-level systems, enabling spin qubit-based quantum information processing \cite{Loss1998, Burkard2023}. Access to spin-resolved states, typically confined in semiconducting quantum dots, requires breaking of the time-reversal symmetry to lift the Kramers degeneracy, often achieved via large magnetic fields. The combination of semiconductors and superconductors into hybrid material platforms \cite{Krogstrup2015, Shabani2016} creates unprecedented opportunities for spin manipulation with Andreev bound states (ABSs)---fermionic modes arising in a semiconducting region bounded by superconductors \cite{Beenakker1991, Furusaki1991, Pillet2010, Chang2013, Bretheau2013a, Bretheau2013b,Janvier2015,Hays2018,Nichele2020}.
Previously, resolving spin-split ABSs was attained in large magnetic fields ($\sim 100~\mathrm{mT}$) \cite{Lee2014, VanWoerkom2017} or by integration of ferromagnetic elements \cite{Sau2010, Escribano2021, Vaitiekenas2022}. 
An intriguing route to locally break time-reversal symmetry without the need of these ingredients is via control over the superconducting phase difference, although it normally requires long Josephson junctions (JJs) with strong spin--orbit coupling (SOC) to lift the spin degeneracy \cite{Chtchelkatchev2003, Beri2008, Reynoso2012, Park2017}.
This led to measured level splittings up to $\sim 1~\mathrm{GHz}$ \cite{Tosi2019, Hays2020, Bargerbos2022a} and enabled the realization of Andreev spin qubits \cite{Chtchelkatchev2003, Padurariu2010, Padurariu2012, Park2017, Hays2021, PitaVidal2023}, that leverage the advantages of both superconducting and spin qubit platforms.

Multiterminal JJs with SOC promise considerable advantages as platforms for superconducting spin manipulation \cite{vanHeck2014}: in such devices, large spin splitting may be induced solely by controlling superconducting phase differences, while remaining in the short-junction limit. Concomitantly, ground state fermion parity transitions (i.e., switches between even and odd number of fermions in the superconducting condensate) are expected, marked by zero-energy Andreev level crossings in the spectrum \cite{Beenakker2013}.
Parity engineering is of crucial importance for realizing artificial Kitaev chains \cite{Kitaev2001, Sau2012, Leijnse2012, Dvir2023} and parity-protected qubits \cite{Doucot2012, Brooks2013, Larsen2020, Gyenis2021}.
Unlike their two-terminal counterparts, multiterminal JJs enable parity tuning in the absence of charging energies and external magnetic fields. Recent experiments revealing ABS spectra in phase-controlled three-terminal JJs (3TJJs) \cite{Coraiola2023} provided a first demonstration of higher-dimensional Andreev band structures \cite{Yokoyama2015, Riwar2016, Xie2017}, particularly in the context of Andreev molecules \cite{Pillet2019, Kornich2019, Matsuo2022, Haxell2023}, but the feasibility of spin-resolved ABSs is yet to be established. 

Here we report on spin-degeneracy breaking and parity transitions controlled by superconducting phase differences in a planar 3TJJ with SOC. We achieve large ABS spin splitting ($\sim 40 ~\mu \mathrm{eV}$, corresponding to a frequency in excess of $9~\mathrm{GHz}$) and level crossings at zero energy, consistent with ground state fermion parity transitions. These phenomena are realized in the absence of external magnetic fields or sizable charging energies. In situ spin and parity tuning is enabled by full phase control through integrated flux-bias lines. The spin nature of the splitting is further supported by magnetic field-dependent studies.  Our results demonstrate a new approach for engineering spin and parity degrees of freedom in hybrid quantum devices.

\begin{figure}
	\includegraphics[width=\columnwidth]{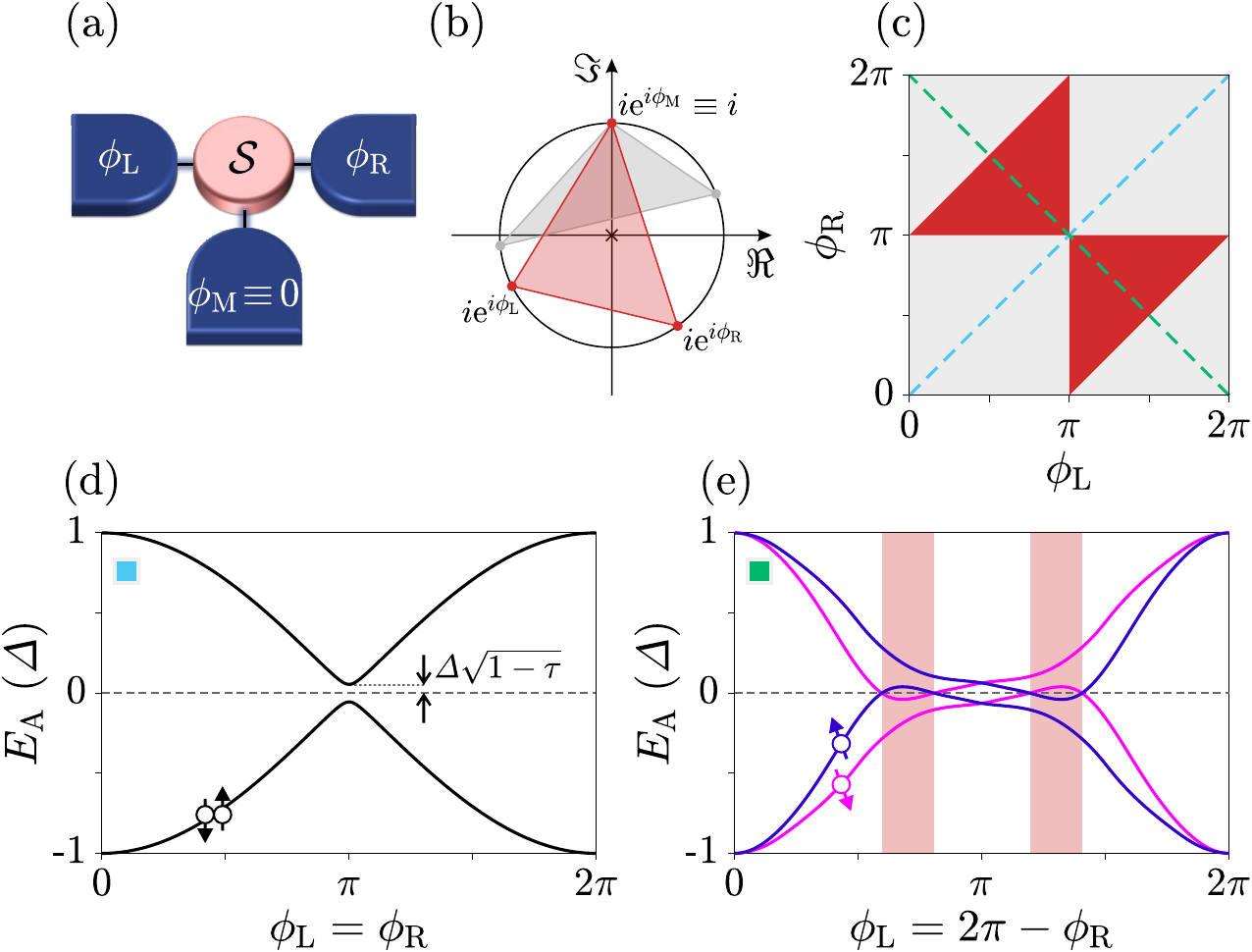}
	\caption{(a) Schematic representation of a hybrid three-terminal Josephson junction (3TJJ). Three superconducting leads with phases $\phiL$, $\phiR$ and $\phiM \equiv 0$ are coupled to a common scattering region $\mathcal{S}$.
		(b) Geometric illustration of the discrete vortex condition, i.e., the necessary condition for the occurrence of zero-energy Andreev bound states (ABSs) and ground state fermion parity transitions. A triangle with vertices $i \mathrm{e}^{i \phi_\alpha}$ (for $\alpha \in \{ \mathrm{L}, \mathrm{R}, \mathrm{M} \}$) is defined on the unitary circle in the complex plane. If it covers the origin (red triangle), the phases wind by $2 \pi$ around the 3TJJ and the discrete vortex condition is fulfilled. Otherwise (gray triangle), zero-energy ABSs and parity transitions are not allowed (adapted from Ref.~\cite{vanHeck2014}).
		(c) Parity diagram in the two-dimensional phase space, spanned by the superconducting phase differences $\phiL$ and $\phiR$. In the red regions, the discrete vortex condition is fulfilled and parity transitions are allowed, whereas parity must stay even in the gray regions. Dashed cyan and green lines indicate the phase-space linecuts $\phiL = \phiR$ and $\phiL = 2 \pi - \phiR$, respectively.
		(d) Energy--phase ABS dispersion along $\phiL = \phiR$, corresponding to a conventional, spin-degenerate ABS described by Eq.~\ref{eq:EA} in the text. Non-perfect transmission $\tau = 0.997$ is assumed, which prevents ABSs from reaching zero energy.
		(e) Dispersion of ABSs along $\phiL = 2 \pi - \phiR$ (schematic representation adapted from Ref.~\cite{vanHeck2014}). Spin degeneracy is lifted due to the the combination of superconducting phases and spin--orbit coupling in $\mathcal{S}$. Blue and magenta ABSs have opposite spin character. Zero-energy Andreev level crossings are allowed and designate parity transitions. Odd-parity regions are highlighted by the red shading.
	}
	\label{fig1}
\end{figure}

A schematic representation of a hybrid 3TJJ is displayed in Fig.~\ref{fig1}(a): three superconducting terminals, with phases $\phiL$, $\phiR$ and $\phiM$, are coupled to a normal scattering region $\mathcal{S}$. Due to gauge invariance, we set $\phiM \equiv 0$, hence $\phiL$ and $\phiR$ correspond to the two independent superconducting phase differences. As derived in Ref.~\cite{vanHeck2014} in the short-junction limit, the necessary condition for a zero-energy ABS to exist in the spectrum is that the phases of the terminals wind by $2\pi$ around the junction. This is referred to as the ``discrete vortex condition'' and is geometrically illustrated in Fig.~\ref{fig1}(b). Assigning to each phase $\phi_\alpha$ (where $\alpha \in \{ \mathrm{L}, \mathrm{R}, \mathrm{M} \}$) the point $i \mathrm{e}^{i \phi_\alpha}$ on the complex plane, the condition is fulfilled when the area of the resulting triangle covers the origin (as shown for example by the red triangle). In this case, the ground state fermion parity can switch from even to odd, and a zero-energy Andreev level crossing marks the transition. If the condition is not fulfilled (e.g., gray triangle), the parity of the system remains even, and the ABS energy cannot reach zero but has a finite lower bound.
In the 2D phase space, values of $(\phiL, \phiR)$ for which a discrete vortex is present in the 3TJJ (i.e., odd fermion parity is allowed) describe a pair of triangular regions [red in Fig.~\ref{fig1}(c)]. We note that when any phase difference is zero ($\phiL=0$, $\phiR=0$ or $\phiL=\phiR$), the system behavior is reduced to that of a two-terminal JJ and its parity has to stay even, except at the points where the other phase difference is $\phi = \pi$. However, in a real junction with non-unity transmission $\tau$, ABSs never reach zero energy, thus the gap cannot close and no parity transition may occur. Rather, ABSs described by the energy--phase dispersion relation \cite{Beenakker1991b}
\begin{equation}
	\Ea(\phi) = \pm \Deltait \sqrt{1- \tau \sin ^2 (\phi/2)},
	\label{eq:EA}
\end{equation}
valid in the short-junction limit, have minimum energy $|\Ea(\pi)|=\Deltait \sqrt{1-\tau}$, where $\Deltait$ is the induced superconducting gap. This corresponds to the spectrum plotted in Fig.~\ref{fig1}(d) as a function of $\phi \equiv \phiL = \phiR$ [cyan line in (c)]. Importantly, we remark that ABSs must remain spin degenerate in this configuration, even if spin-rotation symmetry is broken by SOC in the 3TJJ. A different scenario unfolds when $\phiL \neq \phiR$ (both finite): in this case, if SOC is present, the spin degeneracy of the spectrum can be lifted, leading to a large spin splitting up to a significant fraction of $\Deltait$. In addition, ABSs are allowed to cross at zero energy forming extended regions of odd ground state fermion parity \cite{vanHeck2014}. Both effects are visible in the spectrum of Fig.~\ref{fig1}(e), shown along the phase-space linecut $\phiL = 2\pi - \phiR$ [green in (c)].

\begin{figure}
	\includegraphics[width=\columnwidth]{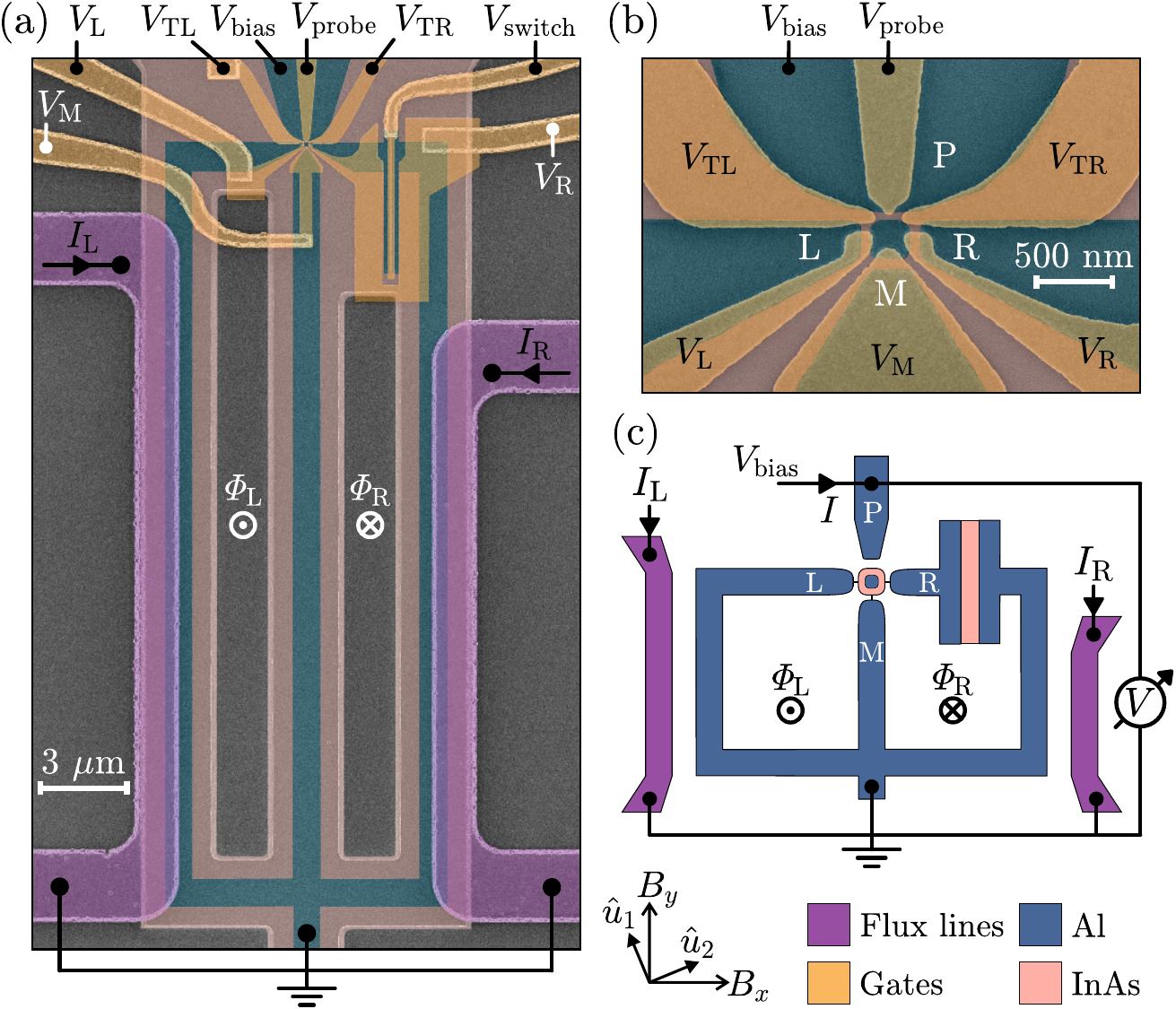}
	\caption{(a) False-colored scanning electron micrograph of a device identical to that under study. Selective removal of Al (blue) exposes the III--V semiconductor below (pink), before being uniformly covered by a dielectric layer (not visible). Gates (yellow) and flux-bias lines (purple) are patterned on top of the dielectric.
		(b) Zoom-in of (a) near the three-terminal Josephson junction (3TJJ) region.
		(c) Schematic of the device together with the measurement setup. Tunneling spectroscopy is performed by measuring the current $I$ flowing through the superconducting probe P (biased by voltage $\Vbias$) and the voltage $V$ across the device with lock-in techniques, to obtain the differential conductance $G$. Independent control over two superconducting phase differences is enabled by the flux-line currents $\Il$ and $\Ir$, that generate external magnetic fluxes $\PhiL$ and $\PhiR$ threading the two superconducting loops. An in-plane magnetic field can be applied with a vector magnet. $B_x$ and $B_y$ directions are indicated, as well as $\hat{u}_2$ and $\hat{u}_1$ that are rotated by $22.5^{\circ}$~counterclockwise.}
	\label{fig2}
\end{figure}

To investigate these phenomena, we realized a 3TJJ in an InAs/Al heterostructure \cite{Shabani2016, Cheah2023}, simultaneously exploiting the strong Rashba SOC in the InAs 2D electron gas (2DEG) and the scalable, top-down patterning approach offered by the heterostructure material. The device, displayed in Fig.~\ref{fig2}, features three superconducting terminals (L, M and R), defined by selective etching of the Al layer and coupled to a common scattering region, forming a 3TJJ. The dimensions of this region, $300~\mathrm{nm} \times 250~\mathrm{nm}$, were chosen to be relatively large on the scale of the spin--orbit length in InAs $l_\mathrm{SO} \sim 150~\mathrm{nm}$ \cite{Fan2020}, yet substantially smaller than the superconducting coherence length of proximitized InAs, $\xi_\mathrm{InAs}\sim600~\mathrm{nm}$. As a result, the effect of SOC is expected to be appreciable in the 3TJJ (unlike the implementation described by Ref.~\cite{Coraiola2023}).
In the middle of the scattering region, a superconducting island of diameter of $200~\mathrm{nm}$ is left for two main purposes: first, it screens stray electric fields and prevents uncontrolled depletion of the 3TJJ from the gates; second, the island facilitates uniform coupling of all terminals to the scattering region, enabling ABSs to depend on all superconducting phases. The small size of the island with respect to the superconducting coherence length prevents the formation of independent two-terminal JJs between each terminal and the island. The three leads L, M and R are connected to a common node defining two closed superconducting loops, which allow for independent tuning of two phases \cite{Coraiola2023}. A JJ integrated on R (referred to as ``switch JJ''), with a length of {40 nm} and a width of {5 $\mu$m}, was designed to have a critical current much larger than between any pairs of L, M and R, therefore the phases of the 3TJJ are not influenced by it. The presence of the switch JJ does not affect the following discussion and a description of its effect is provided in the Supplementary Information. A fourth superconducting terminal (P), biased by the DC voltage $\Vbias$, served as a probe to perform tunneling spectroscopy of sub-gap states in the 3TJJ.
Metallic gate electrodes and flux-bias lines were patterned on top of a dielectric layer, uniformly deposited across the entire sample. Gates energized by the voltages $\Vtl$ and $\Vtr$ were both set to $-1.285~\mathrm{V}$ to deplete the 2DEG below and form a tunneling contact between P and the 3TJJ. In such a weak-coupling regime, the influence of the probe on the rest of the circuit can be neglected and the differential conductance $G$, measured between P and the common node with low-frequency lock-in techniques [see Fig.~\ref{fig2}(c)], represents the density of states (DOS) of the 3TJJ, with a bias shift $\pm \Deltait / e$ introduced by the superconducting probe \cite{Suominen2017, Nichele2020}. Gates covering the normal regions between the island and the three superconducting terminals were kept at $\Vl=\Vm=\Vr=0$ and had the main role of screening the effect of the other gates on the scattering region, without introducing a charging energy. Two additional gates were set to $\Vp=0.42~\mathrm{V}$ and $\Vj=0$ throughout. Currents $\Il$ and $\Ir$ injected into the flux-bias lines generated external magnetic fluxes $\PhiL$ and $\PhiR$ threading the superconducting loops and enabled full control over the 2D phase space. Experiments were performed in a dilution refrigerator with base temperature below $10~\mathrm{mK}$, equipped with a three-axis vector magnet. Magnetic field directions $B_x$ and $B_y$ are indicated in Fig.~\ref{fig2}, together with the additional orientations $\hat{u}_1$ and $\hat{u}_2$ (rotated by $22.5^{\circ}$~counterclockwise with respect to $B_y$ and $B_x$, respectively). Further details regarding materials, fabrication and measurement setup are provided in the Supplementary Information.

\begin{figure}
	\includegraphics[width=\columnwidth]{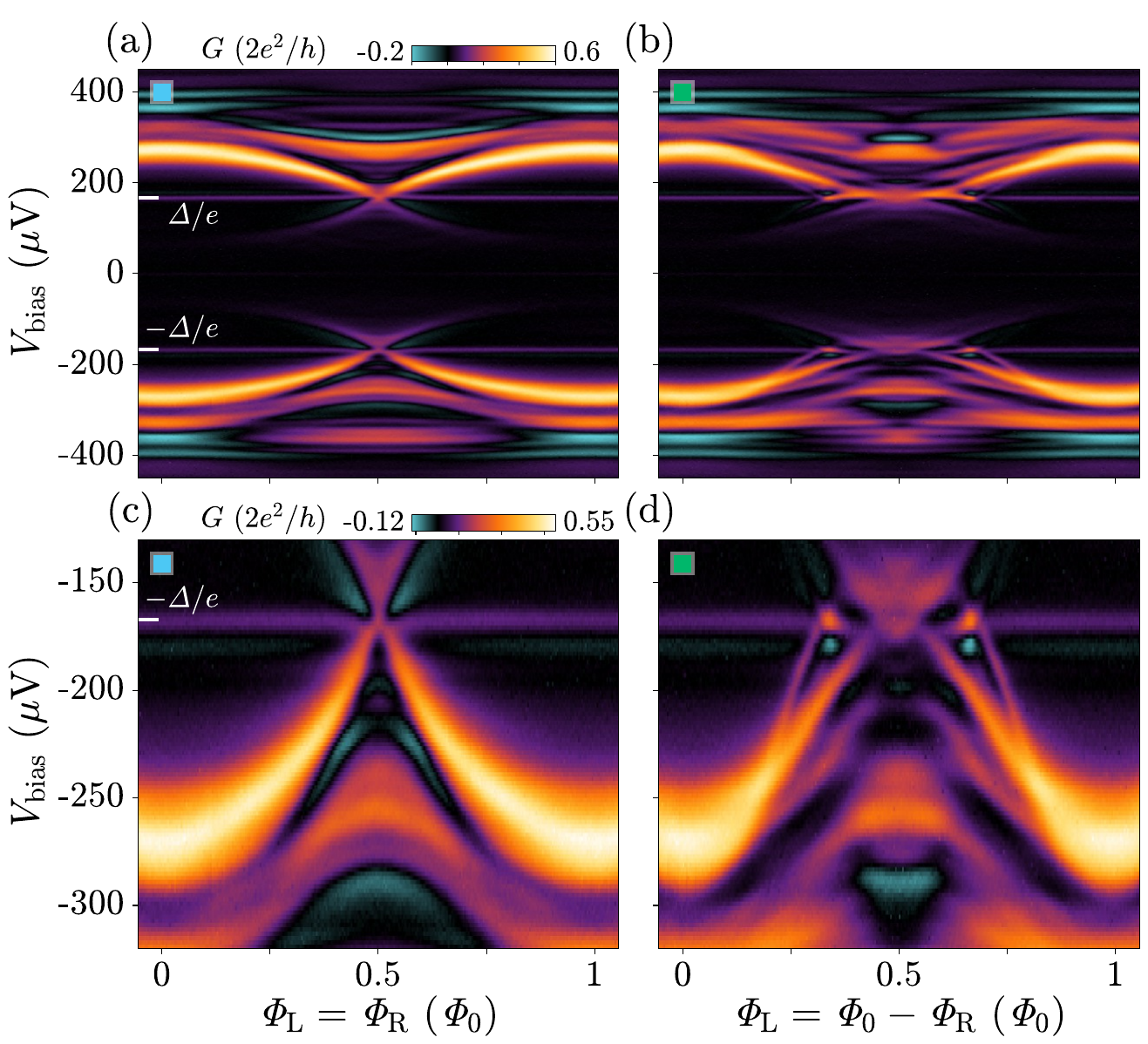}
	\caption{Tunneling spectroscopy of Andreev bound states (ABSs) along phase-space linecuts. (a) Differential tunneling conductance $G$ as a function of voltage bias $\Vbias$ along linecut $\PhiL=\PhiR$, revealing a high-transmission ABS with conventional phase dispersion. Due to the superconducting probe, a transport gap between $\Vbias = \pm \Deltait / e = \pm 167~\mathrm{\mu V}$ is present in the spectrum, delimited by conductance peaks (approximately phase-independent) attributed to multiple Andreev reflection.
		(b) $G$ as a function of $\Vbias$ along linecut $\PhiL = \Phio - \PhiR$. The ABS spectrum is strongly modified compared to (a), showing level splitting, compatible with spin-degeneracy breaking, and zero-energy crossings (highlighted by the conductance enhancement at $|\Vbias| = \Deltait /e$) marking ground state parity transitions in the spectrum.
		(c),(d) Zoom-in of (a),(b) on the ABS spectrum at negative $\Vbias$.}
	\label{fig3}
\end{figure}

\begin{figure}
	\includegraphics[width=\columnwidth]{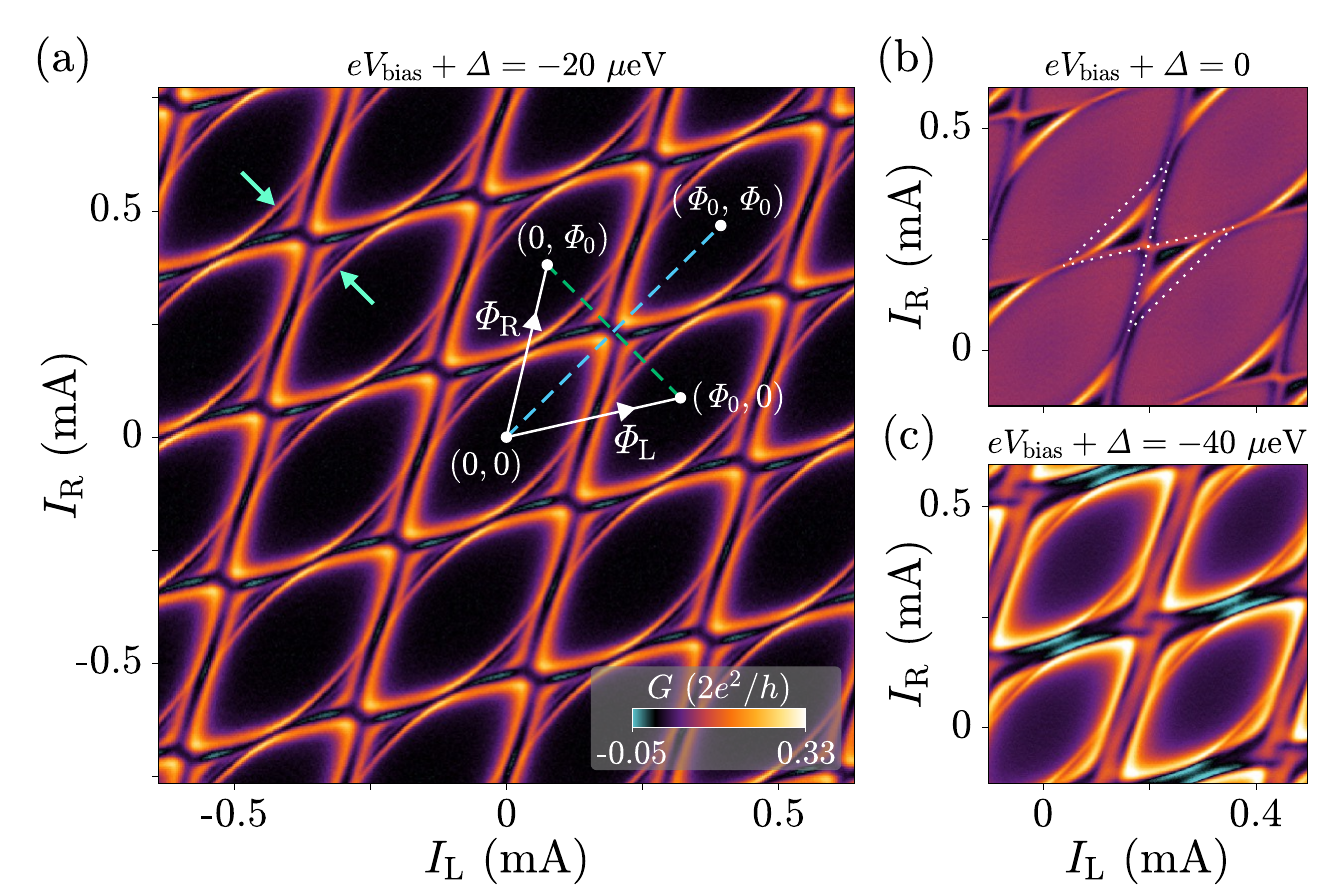}
	\caption{Constant-energy cut planes of the Andreev band structure measured in the two-dimensional phase space.
		(a) Differential tunneling conductance $G$ as a function of flux-line currents $\Il$ and $\Ir$ at fixed voltage bias $\Vbias$, such that $e \Vbias + \Deltait = -20~\mu \mathrm{eV}$ (with $\Deltait = 167~\mathrm{\mu eV}$), resulting in a cut plane $20~\mathrm{\mu eV}$ below the Fermi level. Periodicity along two directions, corresponding to the external magnetic fluxes $\PhiL$ and $\PhiR$, is highlighted by the white segment. Dashed cyan and green lines indicate the phase-space linecuts $\PhiL=\PhiR$ and $\PhiL = \Phio - \PhiR$, shown in Fig.~\ref{fig3}, and correspond to the paths $\phiL = \phiR$ and $\phiL = 2 \pi - \phiR$ plotted in Fig.~\ref{fig1}(c). Split lines marked by the turquoise arrows are related to the spin-split Andreev levels visible in the spectrum [Figs.~\ref{fig3}(b) and \ref{fig3}(d)].
		(b) As (a), but measured at $e \Vbias + \Deltait = 0$ to probe the Andreev band structure at zero energy. Conductance peaks are zero-energy Andreev level crossings in the spectrum, corresponding to transitions in the ground-state parity of the system, and thus enclose regions of odd parity. These are comprised in the phase-space regions where the discrete vortex condition is verified [red in Fig.~\ref{fig1}(c), reported here as the dotted white triangles].
		(c) As (a), but at $e \Vbias + \Deltait = -40~\mu \mathrm{eV}$, showing the Andreev band structure at higher energy.}
	\label{fig4}
\end{figure}

Tunneling spectroscopy measurements of phase-dispersing ABSs are shown in Fig.~\ref{fig3}. The spectra are probed as a function of $\Vbias$ along the two flux-space linecuts $\PhiL=\PhiR$ [panels (a) and (c)] and $\PhiL = \Phio - \PhiR$ [panels (b) and (d)], obtained from linear combinations of the flux-line currents $\Il$ and $\Ir$ [see also Fig.~\ref{fig4}(a)]. Here, $\Phio=h/2e$ is the superconducting flux quantum, with $h$ the Planck constant and $e$ the elementary charge.
These linecuts correspond to those introduced in Figs.~\ref{fig1}(c)--(e) for the 2D phase space (cyan and green line, respectively). All measurements show a transport gap $2 \Deltait / e = 334~\mathrm{\mu V}$ related to the superconducting probe, delimited by conductance peaks at $\Vbias = \pm 167~\mathrm{\mu V}$ with negligible flux dependence and consistent with multiple Andreev reflection occurring at $e |\Vbias| = \Deltait$ \cite{Klapwijk1982}. Above the transport gap, we observe an electron-hole-symmetric spectrum dominated by flux-dependent resonances, representing ABSs in the 3TJJ \cite{Pillet2010, Nichele2020}. Finite conductance features within the transport gap are ascribed to a nonzero DOS in the superconducting probe for energies $|E| \lesssim \Deltait$ \cite{Dynes1978}.
The spectrum shown in Fig.~\ref{fig3}(a) and \ref{fig3}(c) reveals an ABS resembling the conventional energy dispersion of Eq.~\ref{eq:EA} with near-unity transmission, as it forms a sharp cusp that approaches the gap edge very closely. Conversely, the spectrum probed along $\PhiL = \Phio - \PhiR$ [Figs.~\ref{fig3}(b) and \ref{fig3}(d)] exhibits a striking difference from that of Eq.~\ref{eq:EA}: as the ABS moves from high $|\Vbias|$ towards $e|\Vbias|=\Deltait$, it markedly splits into two resonances. Both appear to fully reach $e|\Vbias|=\Deltait$, i.e., the split ABSs cross zero energy. This is more evident for the outermost ABS, that overlaps with $e|\Vbias|=\Deltait$ at a flux significantly different from $\PhiL =0.5 \Phio$, resulting in a point of enhanced conductance; concomitantly, the slope of the state dispersion has a sharp change of sign at the crossing point.

Next, we map the 2D phase space by performing tunneling spectroscopy at fixed values of $\Vbias$ as a function of both $\Il$ and $\Ir$, resulting in constant-energy cut planes of the Andreev band structure. In Fig.~\ref{fig4}(a), where $e \Vbias = -\Deltait - 20~\mathrm{\mu eV} = -187~\mathrm{\mu eV}$, the phase space is scanned over an extended region, showing periodicity along two main directions (white lines) that correspond to the external fluxes $\PhiL$ and $\PhiR$, defined in Fig.~\ref{fig2}. The dashed cyan and green segments represent the $\PhiL=\PhiR$ and $\PhiL=\Phio-\PhiR$ linecuts displayed in Fig.~\ref{fig3}. Two high-transmission ABSs, dispersing with $\PhiL$ and $\PhiR$ respectively, form avoided crossings around the $(\Phio/2 + n \Phio, \Phio/2 + m \Phio)$ points (with $n,m$ integers), which indicates ABS hybridization and the formation of a phase-dependent Andreev molecule \cite{Coraiola2023}. Furthermore, we note additional split lines (turquoise arrows)---only encountered along $\PhiL=\Phio-\PhiR$ and whose paths on the phase space depend on both $\PhiL$ and $\PhiR$---that are the split ABSs described for Figs.~\ref{fig3}(b) and \ref{fig3}(d).
To study zero-energy Andreev level crossings in the phase space, we set $\Vbias = -167~\mathrm{\mu V}$ (i.e., $e \Vbias + \Deltait = 0$), resulting in the cut plane of Fig.~\ref{fig4}(b). Here, we observe conductance resonances defining pairs of triangles, which constitute subregions of the phase space areas fulfilling the discrete vortex condition [red triangles in Fig.~\ref{fig1}(c), whose perimeter is reported as the dotted white triangles in Fig.~\ref{fig4}(b)].
Finally, Fig.~\ref{fig4}(c) shows the cut plane at $e \Vbias + \Deltait = - 40~\mathrm{\mu eV}$, revealing the evolution of the high-transmission ABSs and of the split lines at higher bias, and the occurrence of an additional ABS with lower transmission in the spectrum.

The measurements presented in Fig.~\ref{fig3} are in excellent qualitative agreement with the spectra predicted by Ref.~\cite{vanHeck2014}, schematically summarized in Figs.~\ref{fig1}(d) and \ref{fig1}(e). We thus interpret the ABS splitting as spin degeneracy breaking, and the zero-energy crossings as the indication of a fermion parity transition in the ground state of the system \cite{Beenakker2013}. The two effects, occurring at zero external magnetic field, are a direct manifestation of the Andreev band structure of a 3TJJ with Rashba SOC, and are enabled by control over two superconducting phase differences. Experimentally, such control via flux-bias lines is local and allows for wide tunability of the spin-splitting size and of the parity of the system. 
In our data, we observe a maximum energy splitting of about $40~ \mathrm{\mu eV}$, corresponding to a frequency of over $9~\mathrm{GHz}$ and to $0.24 \Deltait$ (for an induced superconducting gap of $\Deltait = 167~\mathrm{\mu eV}$), that is approximately one order of magnitude larger than what has been achieved in long JJs based on nanowires with SOC \cite{Tosi2019, Hays2021} and very similar to the theoretical prediction \cite{vanHeck2014}.
The cut planes of the Andreev band structure at constant energy (Fig.~\ref{fig4}) show the trajectories of spin-split ABSs and of zero-energy crossings in the 2D phase space. In Fig.~\ref{fig4}(b), the triangles enclosed by the zero-energy conductance peaks represent regions where the ground state fermion parity has transitioned from even to odd. In the following, we will refer to these features as the ``odd-parity triangles''.

\begin{figure}
	\includegraphics[width=\columnwidth]{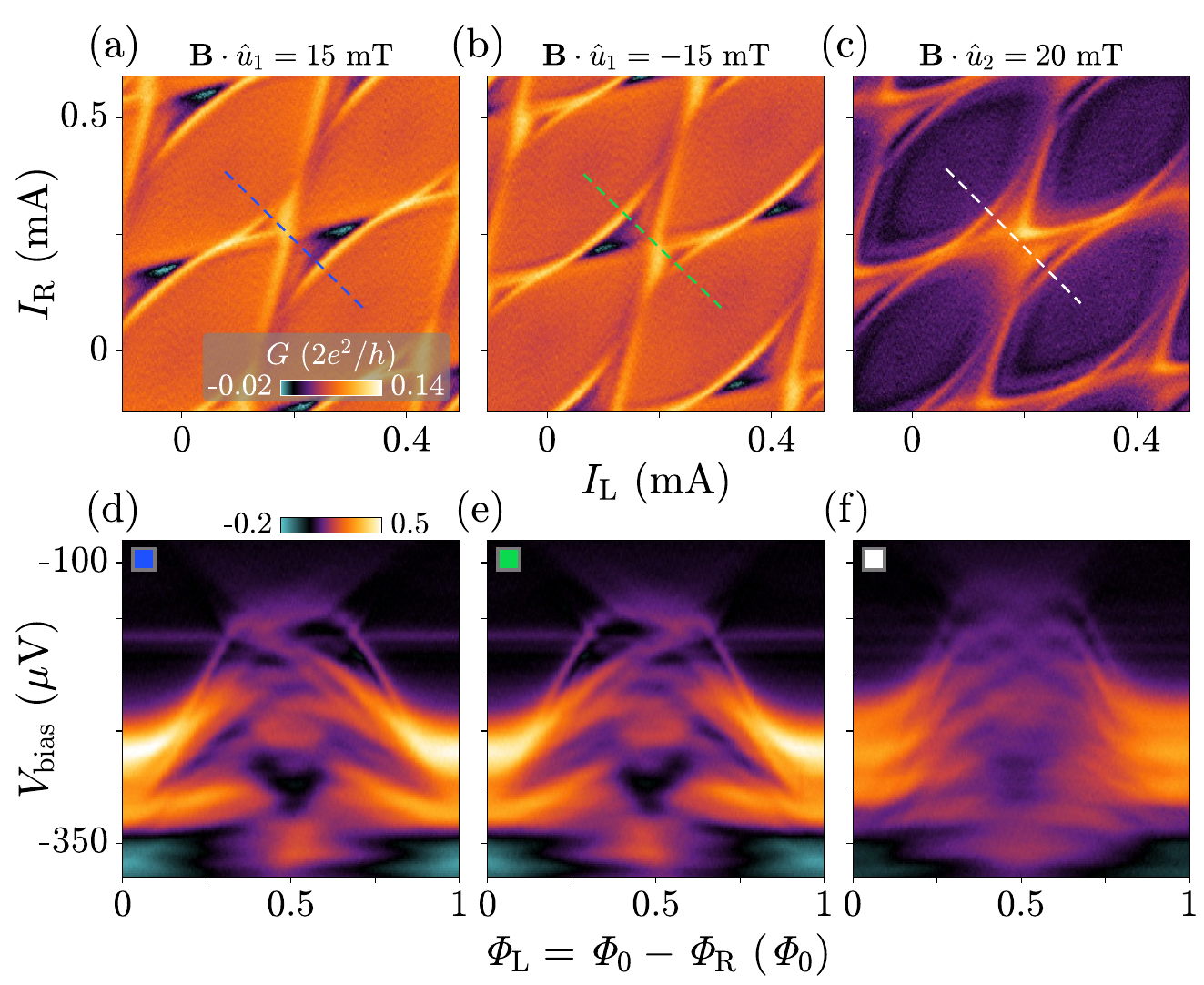}
	\caption{Effect of an in-plane magnetic field on the Andreev bound state (ABS) spectrum.
		(a) Differential tunneling conductance $G$ as a function of flux-line currents $\Il$ and $\Ir$ at fixed voltage bias $\Vbias = -167~\mathrm{\mu V}$, with an in-plane magnetic field $\mathbf{B}$ of magnitude of $15~\mathrm{mT}$ applied along the direction of $\hat{u}_1$ [see Fig.~\ref{fig2}]. The strong asymmetry in the ABS features is attributed to the alignment of $\mathbf{B}$ with the spin--orbit field $\mathbf{B}_\mathrm{SO}$ and indicates that the split lines originate from the spin.
		(b) As (a), but reversing the direction of $\mathbf{B}$, which causes the inversion of all features with respect to $\PhiL = \PhiR = \Phio/2$.
		(c) As (a), but applying a magnetic field of $20~\mathrm{mT}$ along $\hat{u}_2$ (orthogonal to $\hat{u}_1$) and at $\Vbias = -157~\mathrm{\mu V}$. Symmetry of the features is compatible with $\mathbf{B} \perp \mathbf{B}_\mathrm{SO}$.
		(d) Conductance $G$ as a function of $\Vbias$ along the $\PhiL=\Phio-\PhiR$ linecut [dashed blue line in (a)], for a magnetic field of $15~\mathrm{mT}$ along $\hat{u}_1$, revealing the asymmetry in the ABS spectrum.
		(e) As (d), but reversing $\mathbf{B}$. Spectral ABS features are reversed.
		(f) As (d), but with a magnetic field of $20~\mathrm{mT}$ applied along $\hat{u}_2$. The spectrum symmetry is consistent with the cut plane in (c).}
	\label{fig5}
\end{figure}

To further confirm that the origin of the splitting observed in Figs.~\ref{fig3} and \ref{fig4} is related to the spin degree of freedom, we measure additional constant-energy planes and ABS spectra along the $\PhiL=\Phio-\PhiR$ direction while an external in-plane magnetic field $\mathbf{B}$ is applied to the device. Rashba SOC in InAs is expected to result in an in-plane spin--orbit field $\BSO$ perpendicular to the direction of motion \cite{Ihn2010}; however, since the geometry of the 3TJJ does not impose a preferential direction for transport, we cannot make assumptions regarding the orientation of $\BSO$. Instead, we perform a coarse angle dependence of $\mathbf{B}$ and identify directions $\hat{u}_1$ and $\hat{u}_2$ that are inferred to be approximately parallel and orthogonal to $\BSO$ based on the symmetry of the resulting spectrum. When $\mathbf{B} \parallel \hat{u}_1$ [Figs.~\ref{fig5}(a) and \ref{fig5}(d)], we observe a pronounced asymmetry in both the odd-parity triangles [panel (a)] and the spin-split ABS spectrum [panel (d)], as expected in the case that $\mathbf{B}$ has a large component along $\BSO$ \cite{Reynoso2012, Yokoyama2014, Tosi2019}. The spin splitting is enhanced on one side compared to the zero-field case. We also verify that all spectral features flip about $\PhiL = \PhiR = \Phio/2$ upon inversion of the magnetic field direction, as shown in Figs.~\ref{fig5}(b) and \ref{fig5}(e). Conversely, an asymmetry is not expected when the field is applied orthogonal to $\BSO$, which is approximately the situation presented in Figs.~\ref{fig5}(c) and \ref{fig5}(f) for $\mathbf{B} \parallel \hat{u}_2$. 

In summary, we investigated ABS spectra in a phase-controlled 3TJJ with SOC via tunneling spectroscopy experiments. Independent control over two superconducting phase differences enabled access to the Andreev band structure in the 2D phase space. We observed ABS splitting, compatible with breaking of the spin degeneracy, and zero-energy Andreev level crossings, indicating ground state fermion parity transitions, all consistent with theoretical predictions \cite{vanHeck2014}. In the phase space probed at zero energy, odd-parity regions were enclosed by contours of conductance peaks forming pairs of triangles, fulfilling the necessary condition for parity transitions, namely the presence of a discrete vortex in the 3TJJ. Both effects---spin splitting and parity transitions---arise in the absence of external magnetic fields or sizable charging energies, and are tuned by controlling superconducting phase differences. The size of the observed splitting (up to $\sim 40 ~\mathrm{\mu eV}$, or over $9~\mathrm{GHz}$) is substantially larger than previously reported values in nanowire-based long JJs. These results demonstrate that spin and parity degrees of freedom of quantum states are widely controllable by phase tuning in multiterminal hybrid nanostructures. An immediate application is the realization of superconducting spin qubits benefiting from a large, phase-tunable spin-level splitting at zero external fields. Parity control in open systems could be exploited for the engineering of topologically protected Kitaev chains and parity-protected qubits.
The access to transition frequencies between spin-resolved levels of several gigahertz makes our devices ideal for integration with circuit quantum electrodynamics architectures, while in situ and fast frequency tuning via integrated flux-bias lines will potentially enable high-performance logic and multi-qubit coupling schemes.

\section*{Acknowledgements}
We acknowledge fruitful discussions with W.~Belzig, J.~C.~Cuevas, J.~Klinovaja, H.~Legg, D.~Loss, A.~E.~Svetogorov and H.~Weisbrich.
We are grateful to B.~van Heck for useful comments on the manuscript.
We thank the Cleanroom Operations Team of the Binnig and Rohrer Nanotechnology Center (BRNC) for their help and support.
W.W.~acknowledges support from the Swiss National Science Foundation (grant number 200020\_207538).
F.N.~acknowledges support from the European Research Council (grant number 804273) and the Swiss National Science Foundation (grant number 200021\_201082).

\section*{Author contributions}
F.N.~conceived the experiment.
E.C., F.K., R.S.~and W.W.~developed and provided the heterostructure material.
Devices were designed by M.C.~and fabricated by M.C.~and D.Z.H.
M.C.~and F.N.~performed the electrical measurements and interpreted the data, with inputs from D.Z.H., D.S., M.H.~and S.C.t.K.
M.C.~wrote the manuscript with feedback from all of the authors.

\section*{Data availability}
Data presented in this work will be available on Zenodo. The data that support the findings of this study are available upon reasonable request from the corresponding author.

\bibliography{Bibliography}

\clearpage
\newpage
\onecolumngrid

\newcounter{myc}
\renewcommand{\thefigure}{S.\arabic{myc}}

\section*{Supplementary Material: Spin-degeneracy breaking and parity transitions in three-terminal Josephson junctions}

\section{Materials and methods}

Devices were fabricated in a III--V heterostructure grown with molecular beam epitaxy techniques on an InP (001) substrate. The stack consisted of the following layers (starting from the substrate): a 1100 nm thick step-graded InAlAs buffer layer, a 6 nm thick $\mathrm{In_{0.75}Ga_{0.25}As}$ layer, an 8 nm thick InAs layer, a 13 nm thick $\mathrm{In_{0.75}Ga_{0.25}As}$ layer, two monolayers of GaAs and a 15 nm thick Al layer. The latter was deposited in situ without breaking vacuum. The InAs layer hosted a 2DEG that was characterized via measurements in a Hall bar geometry, yielding a peak mobility of  $18000~\mathrm{cm^{2}V^{-1}s^{-1}}$ at an electron sheet density of ${8\cdot10^{11}~\mathrm{cm^{-2}}}$. This resulted in an electron mean free path $l_{e}\gtrsim260~\mathrm{nm}$ and a superconducting coherence length of the 2DEG proximitized by the Al sheet of {$\xi_\mathrm{InAs} = \sqrt{\hbar v_\mathrm{F} l_{e}/\left(\pi \Deltait\right)} \sim 600~\mathrm{nm}$}, with $v_\mathrm{F}$ the Fermi velocity in the 2DEG and $\Deltait$ the induced superconducting gap.

First, large mesa structures were isolated, suppressing parallel conduction between devices and across the middle regions of the superconducting loops. This was done by selectively etching the Al layer with Transene type D, followed by a second chemical etch to a depth of $\sim 380 ~ \mathrm{nm}$ into the III--V material stack, using a $220:55:3:3$ solution of $\mathrm{H_{2}O:C_{6}H_{8}O_{7}:H_{3}PO_{4}:H_{2}O_{2}}$. Next, Al was defined by wet etching with Transene type D at $50\mathrm{^{\circ}C}$ for $4~\mathrm{s}$.
The dielectric, deposited on the entire chip by atomic layer deposition, consisted of a $3~\mathrm{nm}$ thick layer of $\mathrm{Al_2 O_3}$ and a $15~\mathrm{nm}$ thick layer of $\mathrm{HfO_{2}}$. Gate electrodes and flux-bias lines were defined by evaporation and lift-off. In a first step, $5~\mathrm{nm}$ of Ti and $20~\mathrm{nm}$ of Au were deposited to realize the fine features of the gates; in a second, a stack of Ti/Al/Ti/Au with thicknesses $5~\mathrm{nm}$, $340~\mathrm{nm}$, $5~\mathrm{nm}$ and $100~\mathrm{nm}$ was deposited to connect the mesa structure to the bonding pads and to define the flux-bias lines.

Experiments were performed in a dilution refrigerator with base temperature at the mixing chamber below $10~\mathrm{mK}$. 
The sample was mounted on a QDevil QBoard sample holder system, without employing any light-tight enclosure. Electrical contacts to the devices, excepts for the flux-bias lines, were provided via a resistive loom with QDevil RF and RC low-pass filters at the mixing chamber stage, and RC low-pass filters integrated on the QBoard sample holder. Currents in the flux-bias lines were injected via a superconducting loom with only QDevil RF filters at the mixing chamber stage. Signals were applied to all gates and flux-bias lines via home-made RC filters at room temperature.
Electrical measurements were performed with low-frequency AC lock-in techniques, applying a fixed AC voltage $\delta \Vbias = 3~\mu \mathrm{V}$ at frequency $211~\mathrm{Hz}$ to a contact at the superconducting probe (labeled P in Fig.~2 of the Main Text). The DC voltage bias $\Vbias$ was applied differentially: $\Vbias/2$ at the superconducting probe (together with the AC signal) and $-\Vbias/2$ at the common node [indicated as ground in the simplified schematics of Figs.~2(a) and 2(c)]. The latter voltage was provided to the node via a current-to-voltage (I--V) converter, which also measured the AC current $\delta I$ and the DC current $I$ flowing through the device. The four-terminal AC voltage $\delta V$ between the probe and the common node was measured to determine the differential conductance $G \equiv \delta I / \delta V$.

In-plane magnetic fields were applied via a superconducting vector magnet. For each field configuration (including zero field), a finite out-of-plane field component, due to offsets and hysteresis of the magnet as well as to a small misalignment of the sample axes with the magnet axes, had to be compensated. For that, the out-of-plane field was swept across a range $\geq \pm 1~\mathrm{mT}$ and the position of zero field was determined from spectral features such as the size of the superconducting gap and the sharpness of the conductance peaks (subgap states, multiple Andreev reflection peaks and high bias features). Small residual deviations from zero field, caused by hysteresis and imperfect compensation, were taken into account by offsetting the flux-bias line currents $\Il$ and $\Ir$ in such a manner that the point where $\Il = \Ir = 0$ coincided with $\PhiL = \PhiR = 0$ in the constant-bias maps.

\section{Tunnel-gate voltage dependence of differential conductance}

\begin{figure}[h]
	\includegraphics[width=0.5\columnwidth]{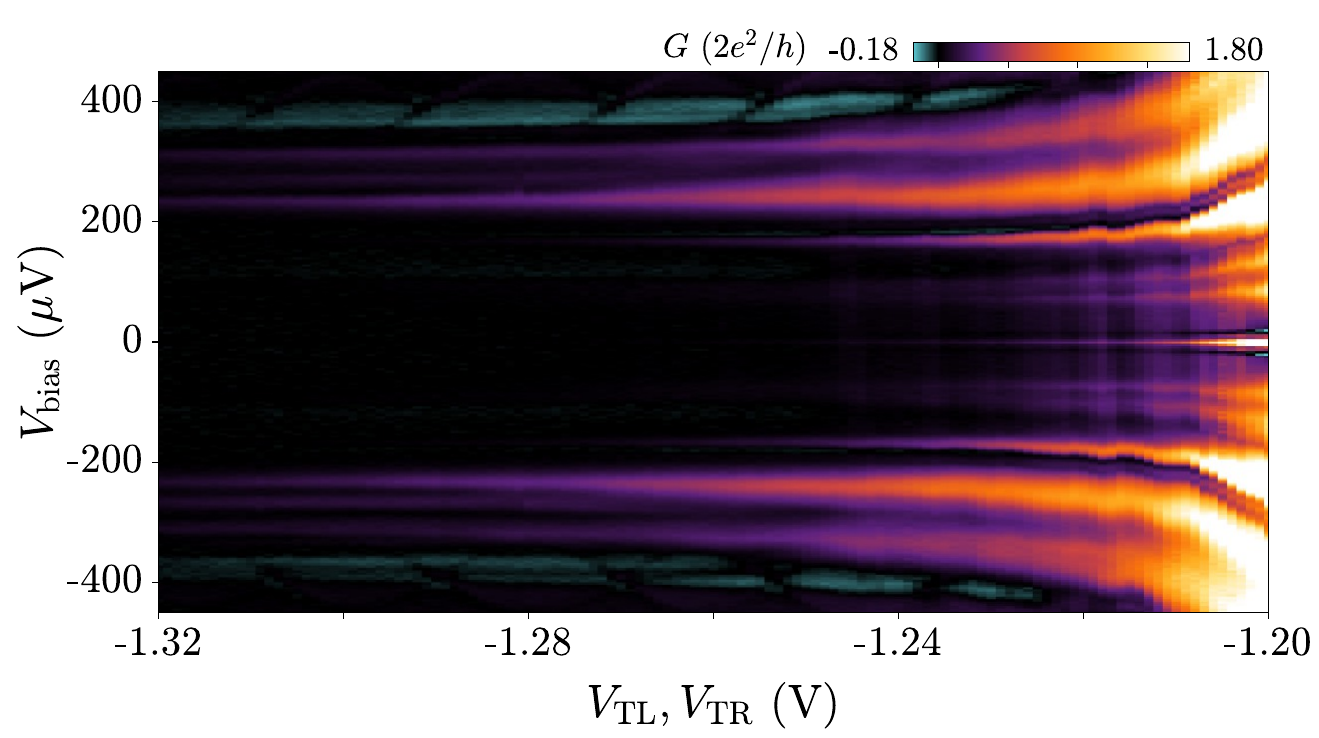}
	\caption{\textbf{Tunnel-gate voltage dependence.} 
		Differential conductance $G$ as a function of voltage bias $\Vbias$ and tunnel-gate voltages $\Vtl = \Vtr$, for $\Vp = 0.42~\mathrm{V}$ and $\Vl = \Vm = \Vr = 0$.}
	\label{figS1}
\end{figure}

The gate electrodes energized by voltages $\Vtl$ and $\Vtr$, denoted as tunnel gates (see Fig.~2 of the Main Text), allowed for electrostatic tuning of the transmission between the superconducting probe and the three-terminal Josephson junction (3TJJ). These gates were always set to a common voltage, $\Vt \equiv \Vtl = \Vtr$. A gate covering the probe (called probe gate), slightly overlapping with the semiconducting region between the tunnel gates, was set to $\Vp = 0.42~\mathrm{V}$ throughout. The three gates surrounding the 3TJJ (labeled left, middle and right gate) and the gate covering the superconductor--semiconductor--superconductor Josephson junction (JJ) on the right arm of the superconducting loops (switch gate) were set to $\Vl = \Vm = \Vr = \Vj = 0$, unless stated otherwise. Figure \ref{figS1} shows the dependence on tunnel gate voltages of the differential conductance $G$ measured as a function of the probe bias voltage $\Vbias$. The conductance decreased as $\Vt$ was reduced, indicating a lower transmission between the probe and the 3TJJ, and a gradual transition from an open to a tunneling regime. In the open regime, conductance peaks were present at zero bias (remnant of a supercurrent between the probe and the leads of the 3TJJ) and at finite bias (related to multiple Andreev reflection (MAR) processes). The tunneling regime is the relevant one to perform tunneling spectroscopy ($\Vt = -1.285~\mathrm{V}$ for the data of the Main Text), as the weak coupling allows for noninvasive probing of the density of states of the 3TJJ. Here, the spectrum is characterized by a transport gap of $334~\mathrm{\mu V}$, consistent with a superconducting gap of the Al probe of $\Deltait = 167~\mathrm{\mu eV}$. Outside the probe gap, pronounced conductance resonances correspond to Andreev bound states (ABSs).

\setcounter{myc}{1}

\section{Dependence of Andreev spectrum on individual phases}

\begin{figure}[h]
	\includegraphics[width=0.5\columnwidth]{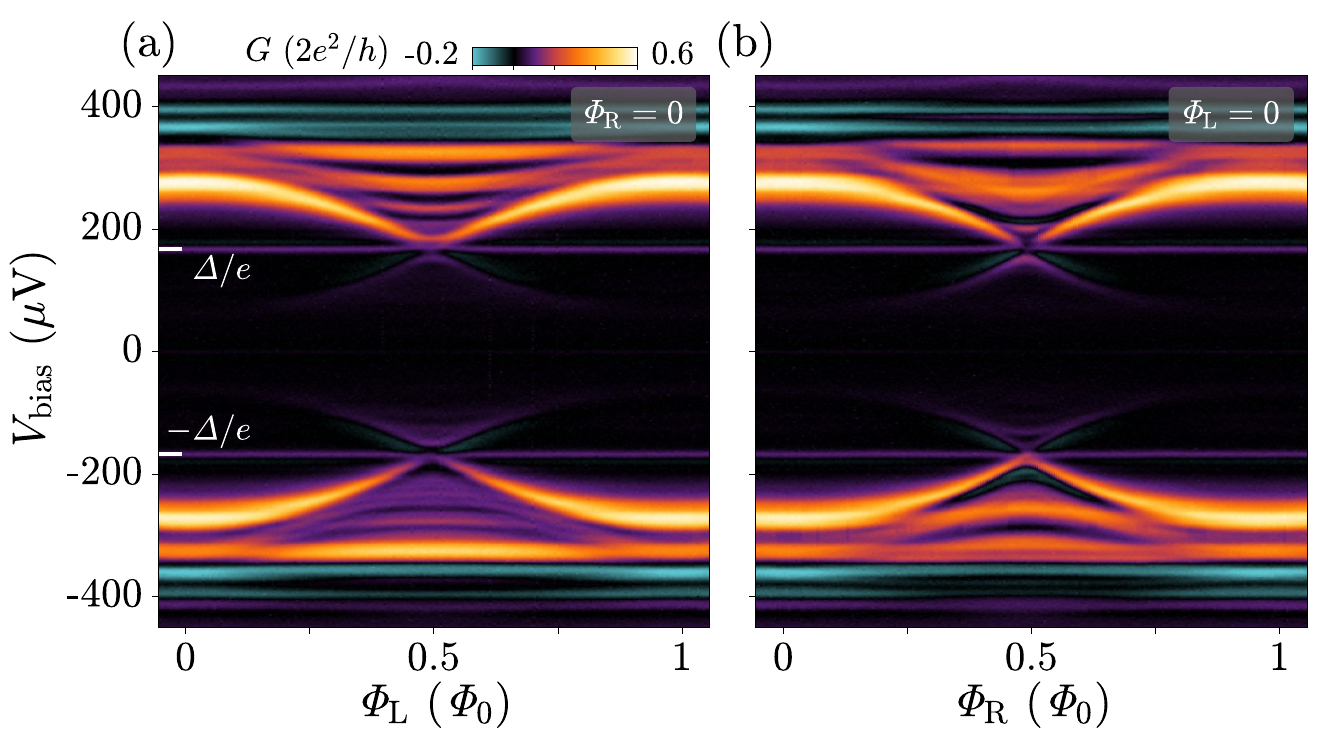}
	\caption{\textbf{Dependence on external magnetic fluxes.} 
		(a) Differential conductance $G$ as a function of voltage bias $\Vbias$ and flux $\PhiL$, for $\PhiR=0$.
		(b) As in (a), but as a function of $\PhiR$ for $\PhiL=0$. Both (a) and (b) correspond to the white linecuts shown in Fig.~4(a) of the Main Text.}
	\label{figS2}
\end{figure}

Tunneling spectroscopy measurements along the phase-space linecuts $\PhiL = \PhiR$ and $\PhiL = \Phio - \PhiR$ (with $\PhiL$ and $\PhiR$ the external magnetic fluxes and $\Phio = h/2e$ the superconducting flux quantum) were presented in Fig.~3 of the Main Text. In this section, we present measurements where each flux was varied individually while keeping the other at zero, which correspond to the white linecuts shown in Fig.~4 of the Main Text. Figures \ref{figS2}(a) and \ref{figS2}(b) show the dependence on $\PhiL$ (for $\PhiR=0$) and on $\PhiR$ (for $\PhiL=0$), respectively, in the same gate configuration as described in the Main Text. Each map reveals a high-transmission ABS, whose energy--phase dispersion resembles the relation $E = \pm \Deltait \left[ 1- \tau \mathrm{sin} ^2 \left( \pi \mathit{\Phi}_\alpha / \Phio \right) \right] ^{1/2}$ valid for short two-terminal JJs \cite{Beenakker1991b}, with $\tau \gtrsim 0.998$ ($\tau$ is the ABS transmission and $\alpha \in \{ \mathrm{L, R} \}$). Here, $\PhiL$- and $\PhiR$-dependent resonances correspond to distinct ABSs, that form in the L--M junction and in the R--M junction regions respectively. As for the $\PhiL = \PhiR$ linecut [Figs.~3(a) and 3(c)], where one phase difference was zero, neither ABS splitting nor signatures of zero-energy crossings are observed, compatible with standard two-terminal JJ physics.

\setcounter{myc}{2}

\section{Voltage bias dependence of constant-energy cut planes}
\begin{figure}[h]
	\includegraphics[width=\columnwidth]{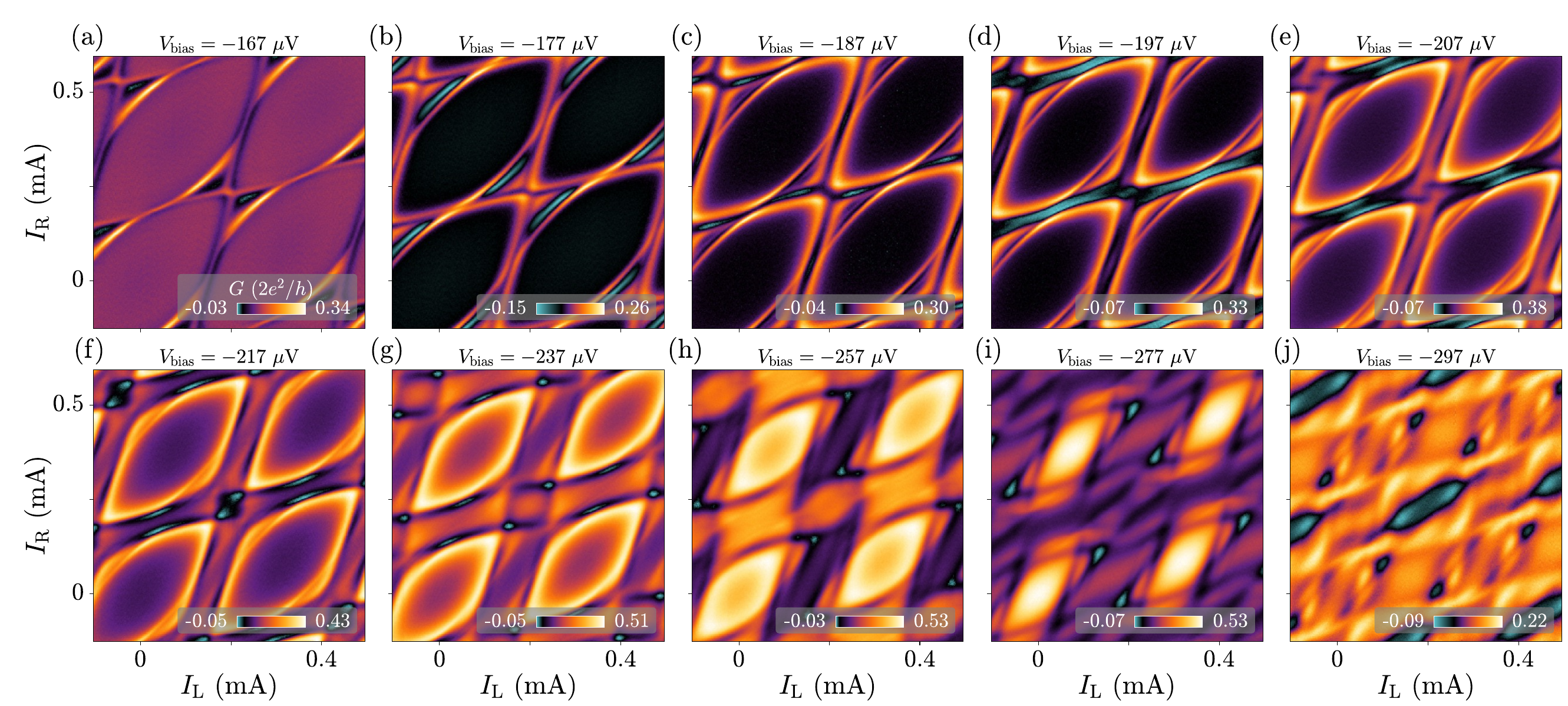}
	\caption{\textbf{Constant-energy cut planes in the two-dimensional phase space for varying voltage bias.}
		(a)--(j) Differential conductance $G$ as a function of flux-line currents $\Il$ and $\Ir$. Each map is at a fixed value of $\Vbias$, ranging from $-167~\mathrm{\mu V} = -\Deltait / e $ (a) through to $-297~\mathrm{\mu V} = -\Deltait / e -130~\mathrm{\mu V}$ (j).} 
	\label{figS3}
\end{figure}

Constant-bias measurements showing the dependence on both flux-line currents---representing cut planes of the Andreev band structure at constant energy---were shown in Fig.~4 of the Main Text. There, selected values of $\Vbias = -167$, $-187$ and $-207~\mu \mathrm{V}$ were presented, corresponding to energies $0$, $-20$ and $-40~\mathrm{\mu eV}$ with respect to the Fermi level (after subtraction of the offset $\Deltait /e = 167~\mathrm{\mu V}$ due to the superconducting probe). In Fig.~\ref{figS3}, we extend the data of Fig.~4 for several values of $\Vbias$ ranging from $-167 ~\mathrm{\mu V}$ (i.e., zero energy) through to $-297~\mathrm{\mu V}$. Each map was measured in the same gate configuration as that of the Main Text. The smaller step size between $\Vbias$ values in the low-energy spectrum allowed us to follow the evolution of the split lines [Figs.~\ref{figS3}(a)--(f)], until they could not be distinguished at higher $|\Vbias|$, compatible with the smaller separation in phase between spin-split ABSs at high energies [see Figs.~1(e) and 3(d) of the Main Text]. Additional $\PhiL$- and $\PhiR$-dependent ABSs with lower transmission first appeared at $\Vbias = -197~\mathrm{\mu eV}$ [Fig.~\ref{figS3}(d)] and $\Vbias = -217~\mathrm{\mu eV}$ [Fig.~\ref{figS3}(f)] respectively, consistent with the spectra displayed in Fig.~\ref{figS2}. The high energy spectrum [for $|\Vbias| \ge 237~\mathrm{\mu V}$, Figs.~\ref{figS3}(g)--(j)] showed a large number of ABSs that could no longer be individually resolved.

\setcounter{myc}{3}

\section{Effect of the switch junction}
\begin{figure}[h]
	\includegraphics[width=0.7692\columnwidth]{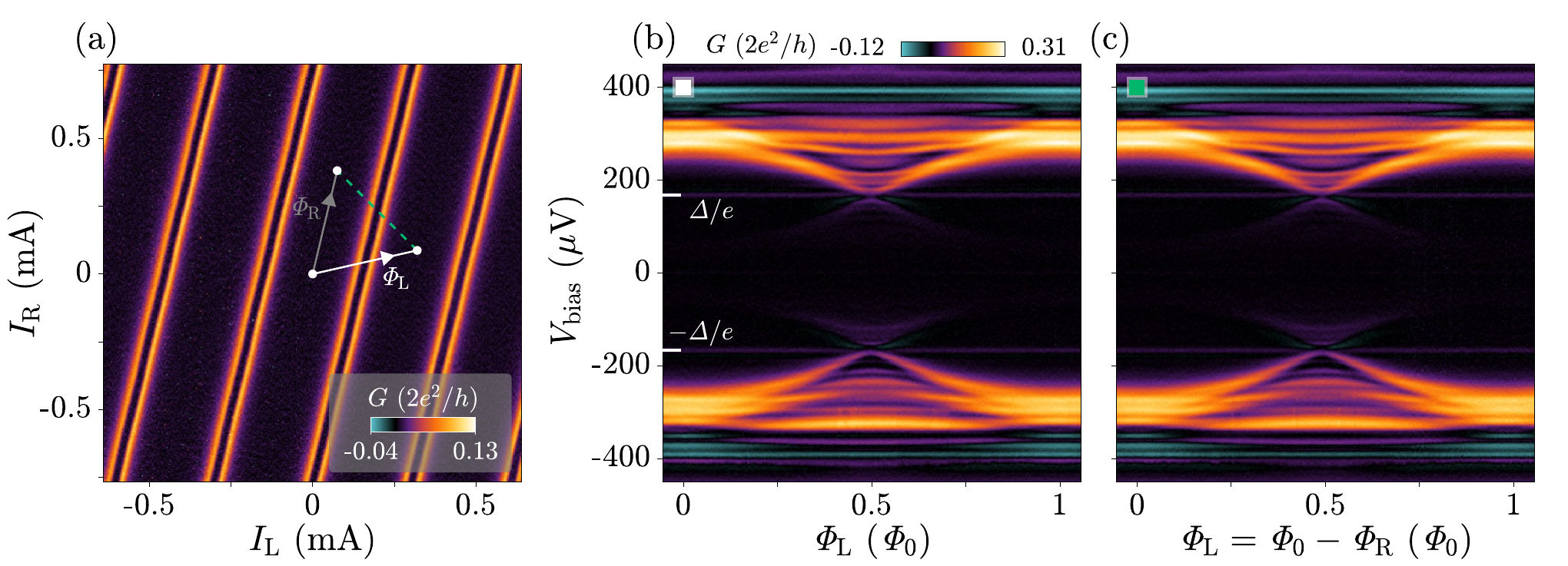}
	\caption{\textbf{Tunneling spectroscopy with a depleted switch junction.}
		(a) Differential conductance $G$ as a function of flux-line currents $\Il$ and $\Ir$ for $\Vj=-2~\mathrm{V}$.
		(b) $G$ as a function of voltage bias $\Vbias$ and external magnetic flux $\PhiL$, namely along the white linecut defined in (a), for $\Vj=-2~\mathrm{V}$.
		(c) As in (b), but along the $\PhiL = \Phio - \PhiR$ linecut [green in (a)].}
	\label{figS4}
\end{figure}

As described in the Main Text, the double-loop geometry features a superconductor--semiconductor--superconductor JJ (labeled switch JJ) on the right superconducting branch (see Fig.~2), that was tunable by the gate voltage $\Vj$. The large critical current of the switch junction makes its inductance small, hence the phase drop across the junction was small and $\PhiL$ and $\PhiR$ only affected the phase differences across the 3TJJ terminals L--M and R--M. Therefore, as long as the switch JJ was not depleted (which was the case with $\Vj=0$, studied so far), its presence could be neglected and the device operated as if both superconducting loops were continuous. Conversely, as shown in Ref.~\cite{Coraiola2023}, depleting the switch junction by setting $\Vj$ to a sufficiently negative voltage interrupted the right loop and made the external flux $\PhiR$ ineffective for tuning the phase difference between terminals R and M. In Fig.~\ref{figS4}, we display the results for $\Vj = -2~\mathrm{V}$, where the switch JJ was completely depleted. The constant-bias cut plane in panel (a), measured at $\Vbias = -187~\mathrm{\mu V}$ as a function of both flux-line currents [as in Fig.~4(a) of the Main Text], confirms that the dependence on $\PhiR$ was suppressed and no features related to the three-terminal geometry were visible. The full spectrum along $\PhiL$, shown in Fig.~\ref{figS4}(b), was similar to that of Fig.~\ref{figS2}(a) (where $\Vj=0$ and $\PhiR = 0$) and revealed again a single high-transmission ABS as well as several states with lower transmission. Importantly, the ABS spectrum was essentially identical when measured along the $\PhiL = \Phio - \PhiR$ linecut [see Fig.~\ref{figS4}(c)], whereas splitting of ABSs and zero-energy conductance peaks was present for $\Vj=0$ [Figs.~3(b) and 3(d) of the Main Text]. These results indicate that such features are specific to the three-terminal geometry with full control over two superconducting phases.
We note that, when the switch JJ was depleted, terminal R was reduced to a floating superconducting island, whose phase was not controlled by external parameters. The low-energy spectrum did not show any hybridization of the $\PhiL$-dependent ABS with the ABS localized between terminal R (floating) and terminal M. This implies that the latter ABS could not enter the low-energy spectrum, namely that the phase difference between R and M always remained sufficiently far from $\pi$.

\setcounter{myc}{4}

\section{Effect of $\Vl$, $\Vm$ and $\Vr$}
The three gates covering the semiconducting regions between each terminal and the superconducting island were so far set to $\Vl = \Vm = \Vr = 0$ and had the main role of screening the effect of the tunnel gates on the 3TJJ, hence preventing its uncontrolled depletion. In this section, we describe the individual and combined effect of finite negative voltages applied to these gates, except for the right gate that was found to be ineffective (voltage $\Vr$ had no influence on any measured quantity). In each gate configuration, the tunnel gate voltages $\Vtl$ and $\Vtr$ were adjusted to keep the transmission of the probe approximately constant (while maintaining $\Vtl = \Vtr$) and to perform spectroscopy in a similar regime as in the Main Text.

\begin{figure}[p!]
	\includegraphics[width=0.5\columnwidth]{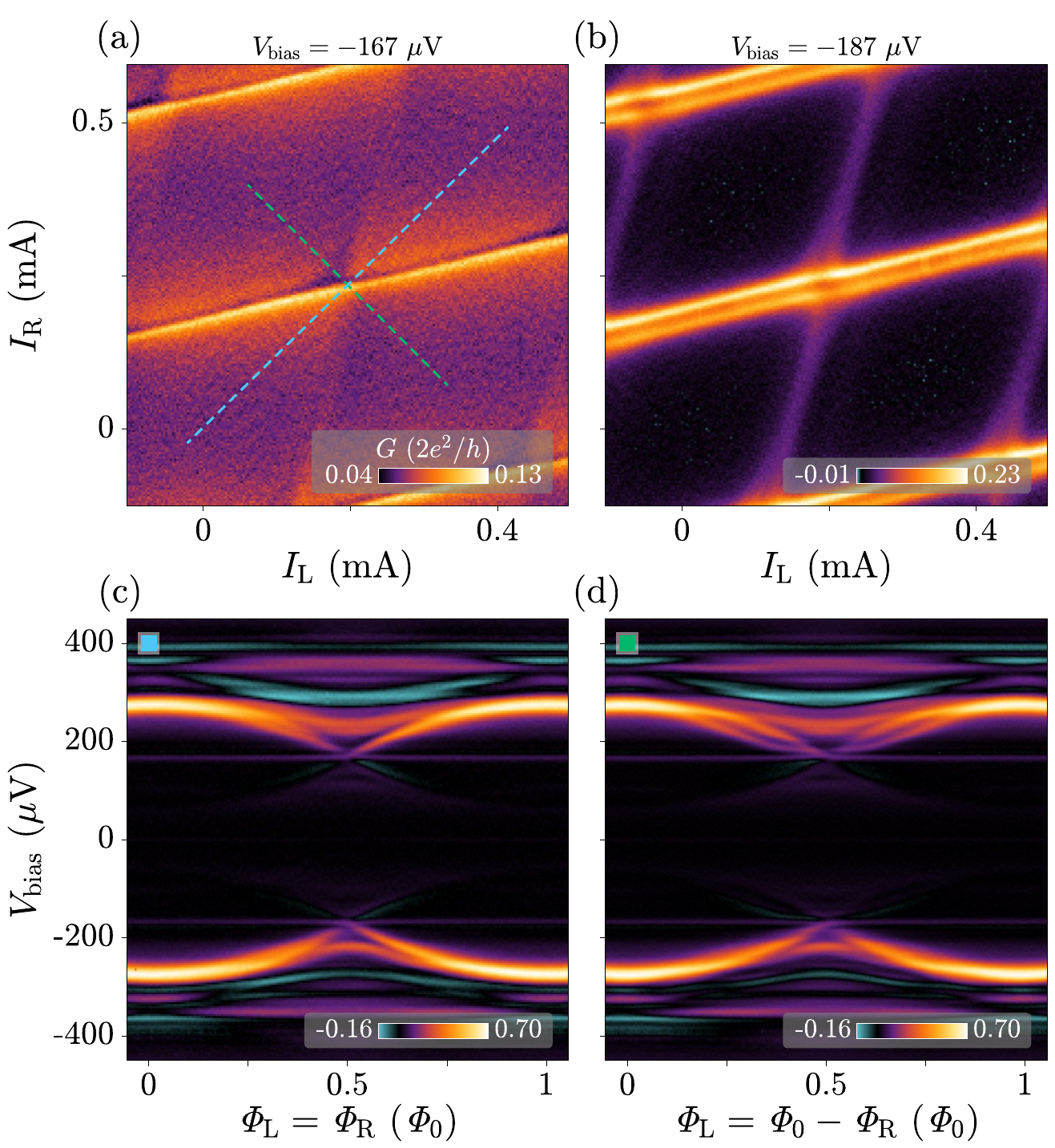}
	\caption{\textbf{Tunneling spectroscopy for $\Vl = -0.4~\mathrm{V}$.}
		(a),(b) Differential conductance $G$ as a function of flux-line currents $\Il$ and $\Ir$, for $\Vbias = -167~\mathrm{\mu V}$ and $\Vbias = -187~\mathrm{\mu V}$, respectively.
		(c),(d) $G$ as a function of $\Vbias$ along the $\PhiL = \PhiR$ linecut [cyan in (a)] and along the $\PhiL = \Phio - \PhiR$ linecut [green in (a)], respectively.}
	\label{figS5}
\end{figure}

For Fig.~\ref{figS5}, we set $\Vl = -0.4~\mathrm{V}$ and adjusted $\Vtl$ and $\Vtr$ to $-1.26~\mathrm{V}$ to deplete the semiconducting region near terminal L and decouple it from the common scattering region. In the constant-bias cut planes [Figs.~\ref{figS5}(a) and \ref{figS5}(b)], the ABS dispersing with $\PhiL$ was substantially less visible and hybridization between ABSs near the intersection points, ABS splitting and odd-parity triangles at zero energy all disappeared. This is confirmed by the phase-space linecuts $\PhiL = \PhiR$ and $\PhiL = \Phio-\PhiR$ displayed in Figs.~\ref{figS5}(c) and \ref{figS5}(d), which revealed similar spectra (unlike in Fig.~3 of the Main Text) characterized by a conventional ABS dispersion.

\setcounter{myc}{5}

\setcounter{myc}{6}
\begin{figure}[p!]
	\includegraphics[width=0.5\columnwidth]{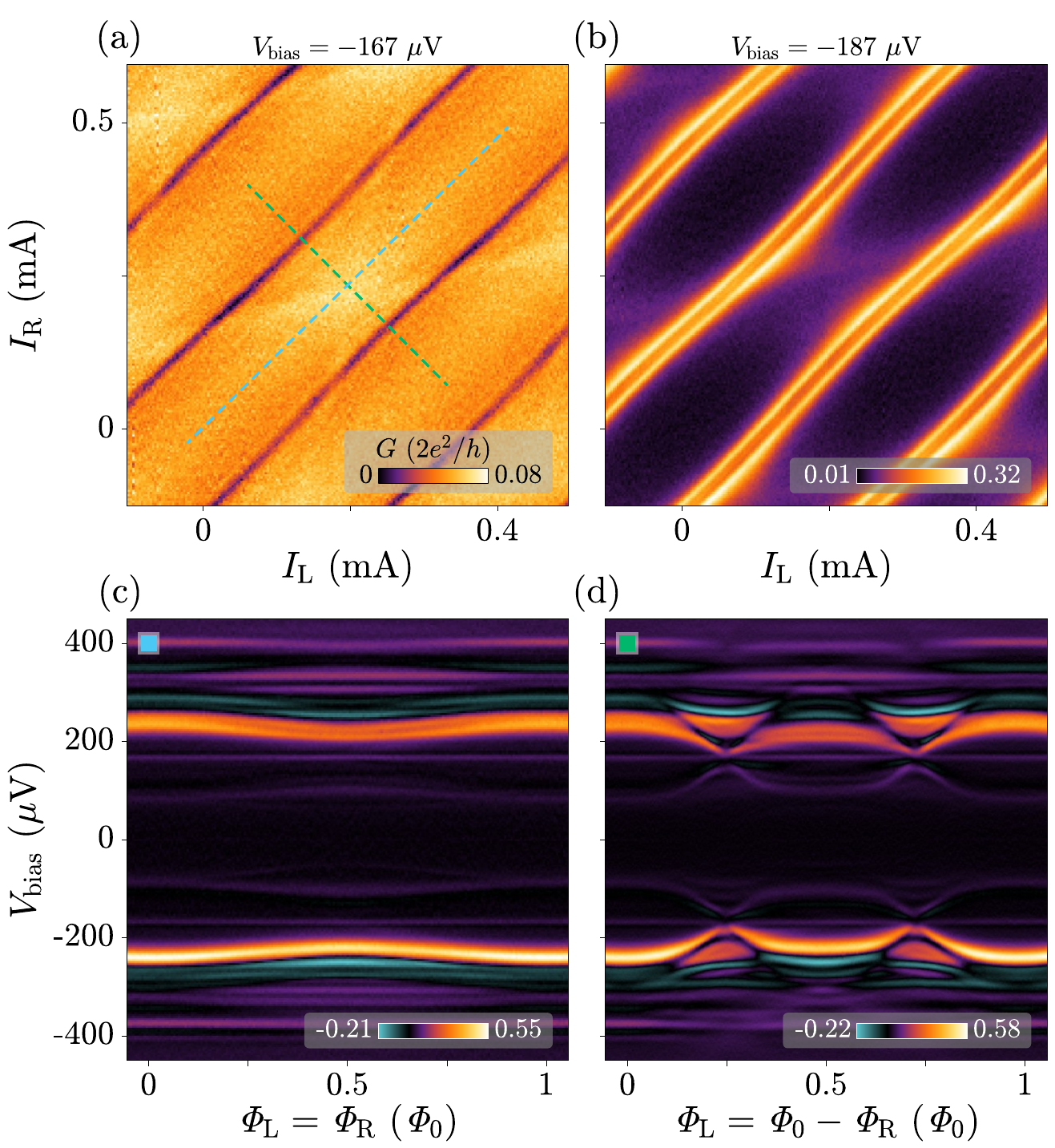}
	\caption{\textbf{Tunneling spectroscopy for $\Vm = -0.6~\mathrm{V}$.}
		(a),(b) Differential conductance $G$ as a function of flux-line currents $\Il$ and $\Ir$, for $\Vbias = -167~\mathrm{\mu V}$ and $\Vbias = -187~\mathrm{\mu V}$, respectively.
		(c),(d) $G$ as a function of $\Vbias$ along the $\PhiL = \PhiR$ linecut [cyan in (a)] and along the $\PhiL = \Phio - \PhiR$ linecut [green in (a)], respectively.}
	\label{figS6}
\end{figure}

Next, we set $\Vl$ back to $0$ and applied a voltage $\Vm = -0.6~\mathrm{V}$ to the middle gate to decouple terminal M from the 3TJJ region. Here, $\Vtl$ and $\Vtr$ were tuned to $-1.252~\mathrm{V}$. In this new gate configuration, we performed the same four measurements as in Fig.~\ref{figS5} and show the result in Fig. \ref{figS6}. Interestingly, both original ABS resonances (one depending only on $\PhiL$, the other only on $\PhiR$) became extremely feeble and spectroscopy was dominated by diagonal features oriented along $\PhiL = \PhiR$. Consequently, these states depend on $\PhiL - \PhiR$, i.e., the total flux threading both superconducting loops (note that positive $\PhiL$ and $\PhiR$ were conventionally defined in opposite directions, see Fig.~2 of the Main Text), and they indicate the presence of a conducting channel between terminals L and R. As this was not observed in previous measurements without a superconducting island in the middle of the normal scattering region \cite{Coraiola2023}, we conclude that the island ensures a more uniform coupling across the three terminals, with a high-transmission channel even between the terminals with the largest separation. From the spectrum of Fig.~\ref{figS6}(d), we also note that no ABS splitting nor conductance enhancement at zero energy occurred, as in Fig.~\ref{figS5} for the left gate depleted. 

Figures \ref{figS4}--\ref{figS6} show that our devices support three types of ABS, depending on $\PhiL$, $\PhiR$ and $\PhiL - \PhiR$ respectively. Each could be visualized individually by decoupling one of the three superconducting terminals. A conducting channel was present between all pairs of terminals, although none of them alone displayed level splitting and crossings at zero energy. These phenomena were exclusively possible in our devices with all three superconductors coupled to the scattering region and via control over both phase differences.

\begin{figure}[h]
	\includegraphics[width=0.7692\columnwidth]{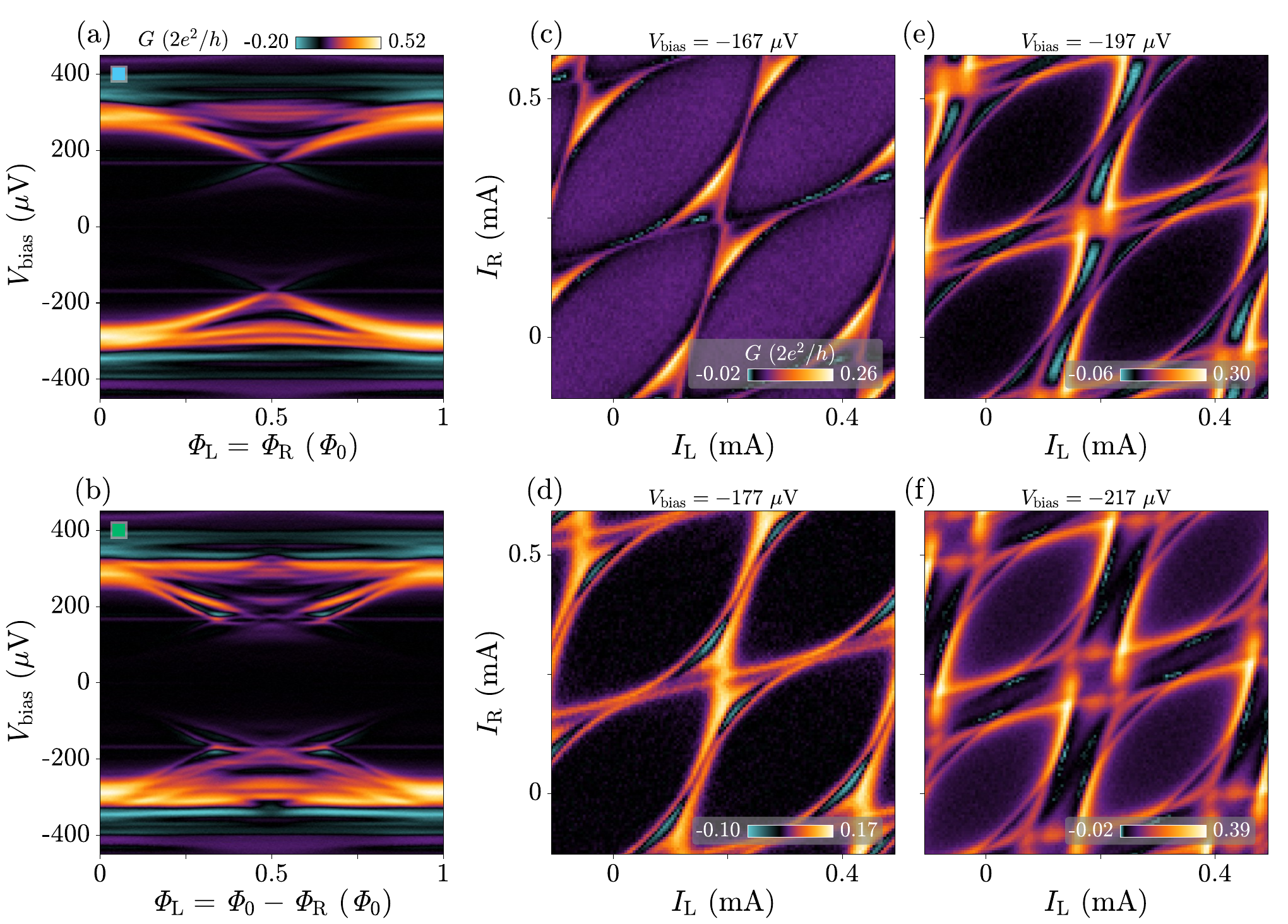}
	\caption{\textbf{Tunneling spectroscopy for $\Vl = \Vm = -0.2~\mathrm{V}$.}
		(a),(b) Differential conductance $G$ as a function of $\Vbias$ along the $\PhiL = \PhiR$ linecut and along the $\PhiL = \Phio - \PhiR$ linecut, respectively.
		(c)--(f) $G$ as a function of flux-line currents $\Il$ and $\Ir$, for $\Vbias$ ranging from $-167~\mathrm{\mu V}$ to $-217~\mathrm{\mu V}$.}
	\label{figS7}
\end{figure}

Finally, we set both $\Vl$ and $\Vm$ to $-0.2~\mathrm{V}$ (adjusting $\Vtl$ and $\Vtr$ to $-1.25~\mathrm{V}$), for which the semiconductor below the left and middle gate was affected but not depleted. Measurements in this new configuration are reported in Fig.~\ref{figS7}: the phase-space linecuts [panels (a) and (b)] correspond to Figs.~3(a) and 3(b) of the Main Text, while the constant-energy cut planes [(c)--(f)] are to be compared to Figs.~4 and \ref{figS3}. The results are qualitatively similar to those observed at $\Vl = \Vm = 0$, indicating that the phenomena discussed in the Main Text were relatively robust and did not depend on the specific gate configuration, nor relied on fine tuning. At relatively large $|\Vbias|$, individual ABSs were more clearly resolved in this configuration than at $\Vl = \Vm = 0$ [compare for example Figs.~\ref{figS7}(f) and \ref{figS4}(f)], which is compatible with a reduced number of low-transmission modes at negative gate voltages.

\setcounter{myc}{7}

\section{Remapping of flux-line-current space to flux space}

\begin{figure}[h]
	\includegraphics[width=0.5\columnwidth]{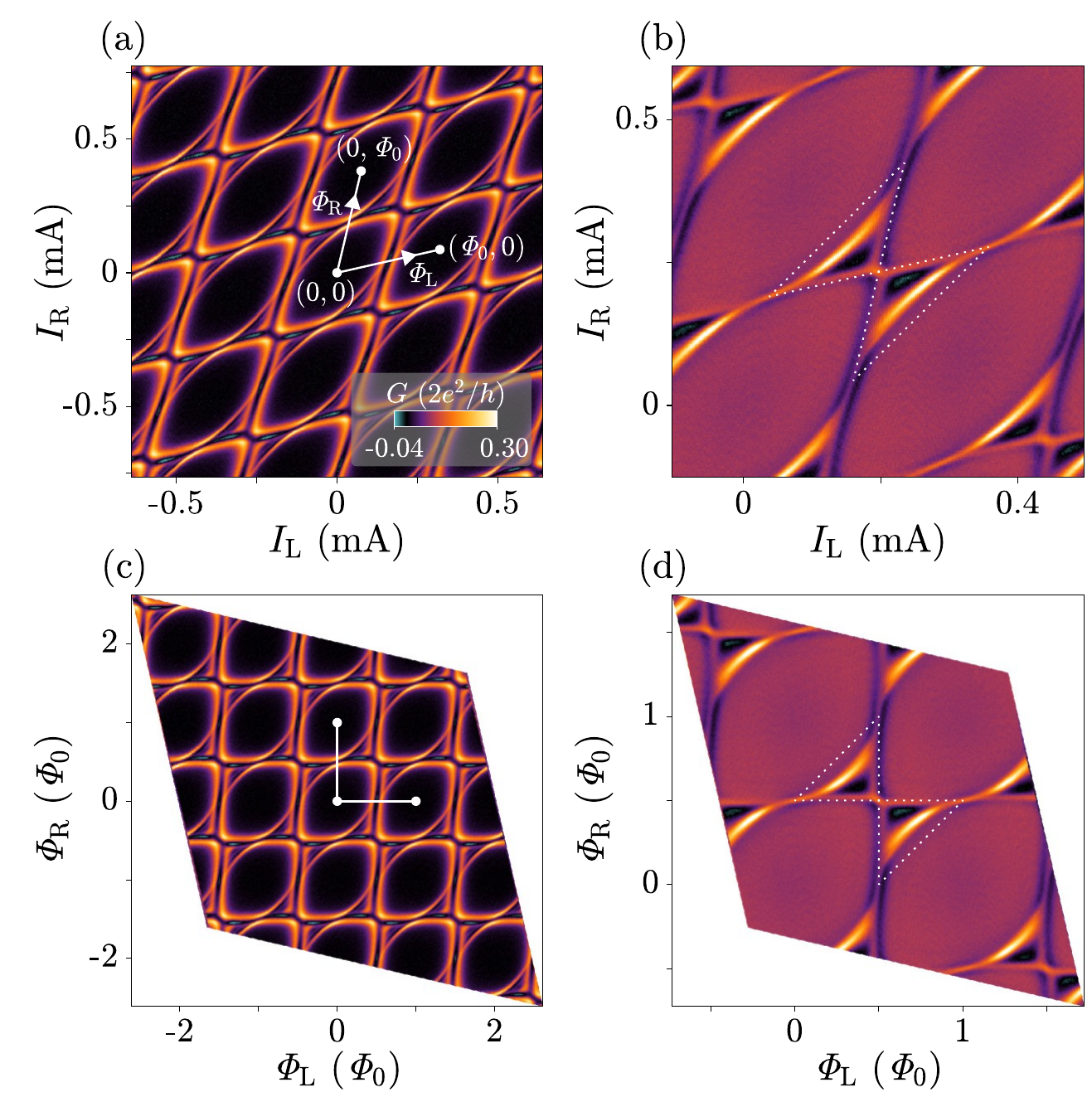}
	\caption{\textbf{Current-to-flux remapping.} 
		(a),(b) Differential conductance $G$ as a function of flux-line currents $\Il$ and $\Ir$, for $\Vbias = -187~\mathrm{\mu V}$ and $\Vbias = -167~\mathrm{\mu V}$, respectively [replotted from Figs.~4(a) and 4(b) of the Main Text].
		(c),(d) As in (a) and (b), but remapping the $(\Il, \Ir)$ axes to the $(\PhiL, \PhiR)$ axes [white lines in (a)], using the linear transformation described in the text.
		In (b) and (d), the dotted white triangles represent the perimeter of the regions where the discrete vortex condition is fulfilled.}
	\label{figS8}
\end{figure}

Full control over two superconducting phase differences was enabled in our devices by superconducting flux-bias lines in combination with the double-loop geometry (see Fig.~2 of the Main Text). Currents $\Il$ and $\Ir$ generated small local magnetic fields that threaded the two superconducting loops, resulting in the external magnetic fluxes $\PhiL$ and $\PhiR$. The left flux-bias line mainly controlled $\PhiL$ due to its proximity to the left loop, however it also produced a smaller flux that threaded the right loop and changed $\PhiR$. An identical argument applies to the right flux line, hence independent flux line control allowed full mapping of the two-dimensional phase space, but required cross-talk between the two lines to be taken into account. This explains the finite slope of the $\PhiL$ and $\PhiR$ axes on the  $(\Il,\Ir)$ plane, as highlighted in Fig.~4(a) of the Main Text [plotted again in Fig.~\ref{figS8}(a)]. Fluxes $\PhiL$ and $\PhiR$ were related to the flux-line currents $\Il$ and $\Ir$ via the linear transformation:
\begin{equation}\label{eqM}
	\begin{pmatrix}
		\PhiL \\
		\PhiR
	\end{pmatrix}
	= \mathbf{M} \cdot 
	\begin{pmatrix}
		\Il \\
		\Ir
	\end{pmatrix}
	=
	\begin{pmatrix}
		M_\mathrm{LL}  & M_\mathrm{LR} \\
		M_\mathrm{RL}  & M_\mathrm{RR}
	\end{pmatrix}
	\cdot
	\begin{pmatrix}
		\Il \\
		\Ir
	\end{pmatrix},
\end{equation}
where $\mathbf{M}$ is the mutual inductance matrix. The matrix $\mathbf{M}$ can be calculated from the constant-bias cut planes knowing both the $(\Il,\Ir)$ coordinates and the $(\PhiL,\PhiR)$ coordinates of two points, as the origin $(0,0)$ is the same in both current space and flux space. For instance, we consider the points $(\mathit{\Phi}_0, 0)$ and $(0, \mathit{\Phi}_0)$ [see Fig.~\ref{figS8}(a)] and substitute their $(\Il,\Ir)$ coordinates in Eq.~\ref{eqM} to obtain a $4 \times 4$ equation system. As a result, we determine the mutual inductance matrix:
\begin{equation}\label{eqM2}
	\mathbf{M} = 
	\begin{pmatrix}
		6.83 ~ \mathrm{pH}  & -1.34 ~ \mathrm{pH} \\
		-1.56 ~ \mathrm{pH} & 5.74 ~ \mathrm{pH}
	\end{pmatrix}
	.
\end{equation}
The ratio $M_\mathrm{RR}/M_\mathrm{LL} \approx M_\mathrm{LR}/M_\mathrm{RL} \approx 0.85$ approximately corresponds to the ratio between the lengths of the vertical segments of the right and left flux line.
While constant-bias cut planes were so far plotted on the $(\Il,\Ir)$ axes to directly visualize the raw data, knowing $\mathbf{M}$ enables a conversion to the $(\PhiL,\PhiR)$ axes using Eq.~\ref{eqM}. This is illustrated in Figs.~\ref{figS8}(c) and \ref{figS8}(d) for the datasets of Figs.~\ref{figS8}(a) and \ref{figS8}(b), respectively. In panels (b) and (d), the dotted white triangles represent the perimeter of the region where the discrete vortex condition is fulfilled.

\setcounter{myc}{8}

\section{Results for additional orientations of the in-plane magnetic field}

\begin{figure}[h]
	\includegraphics[width=\columnwidth]{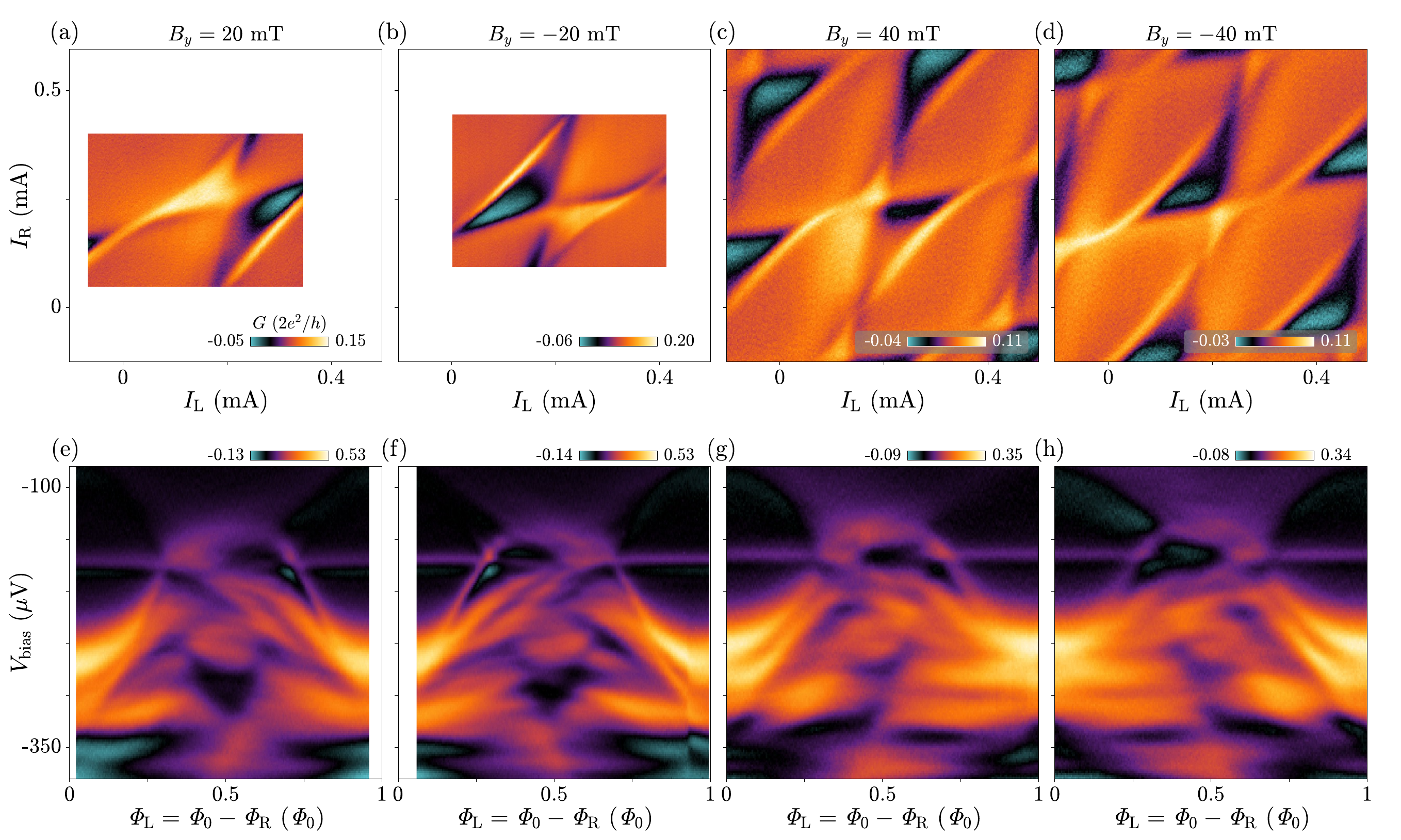}
	\caption{\textbf{Dependence on in-plane magnetic field $B_y$.} 
		(a)--(d) Differential conductance $G$ as a function of flux-line currents $\Il$ and $\Ir$, for $\Vbias = -167~\mathrm{\mu V}$ and $B_y = \pm 20~\mathrm{mT},~\pm 40 ~\mathrm{mT}$.
		(e)--(h) $G$ as a function of $\Vbias$ along the $\PhiL = \Phio - \PhiR$ linecut, at the same $B_y$ as in (a)--(d), respectively.}
	\label{figS9}
\end{figure}

The dependence on an in-plane magnetic field $\mathbf{B}$ was shown in Fig.~5 of the Main Text for the selected orientations $\hat{u}_1$ and $\hat{u}_2$, rotated by $22.5^{\circ}$ counterclockwise with respect to the $x$- and $y$-axes (see Fig.~2) and interpreted to be roughly parallel and orthogonal to the spin--orbit field $\BSO$, respectively. These measurements supported the interpretation that the level splitting observed in the device is a spin effect, particularly related to spin--orbit coupling (SOC). Without prior knowledge about the direction of $\BSO$, we first determined the effect of $B_x$ and $B_y$ on the spectrum and inferred the direction approximately orthogonal to $\BSO$, as described in the following.
The dependence on $B_y$ is summarized in Fig.~\ref{figS9} for fields of 20 and 40 mT applied in both the positive and negative direction. Its effect was qualitatively similar to that of the field applied along $\hat{u}_1$ [Figs.~5(a),(b),(d),(e) of the Main Text]: the odd-parity triangles [Figs.~\ref{figS9}(a)--(d)] and the bias-dependent spectra [Figs.~\ref{figS9}(e)--(h)] became visibly asymmetric, and flipped about $\PhiL=\PhiR=\Phio/2$ upon inversion of the field direction. The effect was more pronounced for larger $B_y$, with the smaller triangle almost completely disappearing. In the spectrum, the splitting was enhanced with respect to the zero-field case, as expected from $\mathbf{B}$ having a large component along $\BSO$.

\setcounter{myc}{9}

Incidentally, we note that the in-plane magnetic field affected the superconducting properties of both the probe and the flux-bias lines. Since the probe was constituted by a thin film of Al and was relatively narrow along the $x$-direction, it was more strongly affected when $\mathbf{B}$ had a large component along $x$ and showed enhanced resilience when $\mathbf{B}$ was largely aligned with the $y$-axis \cite{Suominen2017}. As a consequence, the superconducting probe gap appeared softer along $B_x$ than along $B_y$ (and along $\hat{u}_2$ than along $\hat{u}_1$, as seen in Fig.~5 of the Main Text). Regarding the flux-bias lines, their switching currents (well above $1~\mathrm{mA}$ at $\mathbf{B}=0$) were reduced for increasing fields, with a marginal dependence on the field orientation. For $B_y = \pm 20~\mathrm{mT}$, the lines remained superconducting within $\Il$ and $\Ir$ ranges of about $400$ and $350~\mathrm{\mu A}$ [we recall that the origin of the $(\Il, \Ir)$ planes was shifted to correspond to $\PhiL=\PhiR=0$, as explained in the Methods section]. At the limits of these ranges, flux-line switches to the normal state would have led to discontinuities on the constant-bias cut planes, that were hence measured limiting $\Il$ and $\Ir$ to remain in the superconducting state. Conversely, at $B_y = \pm 40~\mathrm{mT}$ the flux lines were always normal, which did not cause any discontinuity but an aperiodicity in the magnitude of $G$ on the cut planes [visible as regions of enhanced $G$ in Figs.~\ref{figS9}(c),(d) and 5(c)]. This effect is attributed to local heating of the device by the normal lines, that is dependent on $\Il$ and $\Ir$, and did not affect the qualitative behavior of the features observed.

\begin{figure}[h]
	\includegraphics[width=0.5\columnwidth]{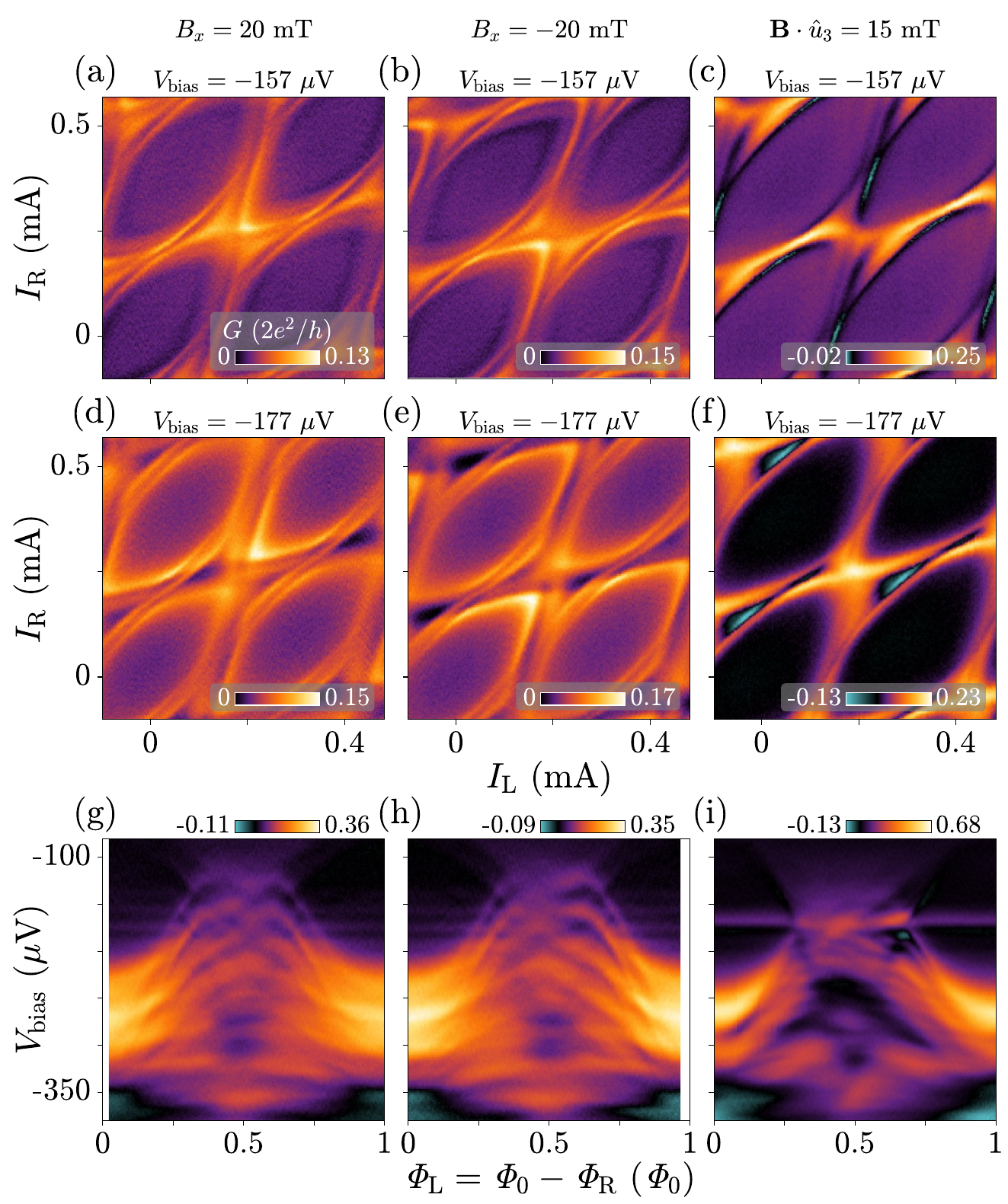}
	\caption{\textbf{Additional in-plane field orientations.} 
		(a)--(f) Differential conductance $G$ as a function of flux-line currents $\Il$ and $\Ir$, for $\Vbias = -157~\mathrm{\mu V}$ [(a)--(c)] and $\Vbias = -177~\mathrm{\mu V}$ [(d)--(f)].
		(g)--(i) $G$ as a function of $\Vbias$ along the $\PhiL = \Phio - \PhiR$ linecut. Each column is characterized by a different in-plane field: $20~\mathrm{mT}$ parallel to the $x$-axis (left), $20~\mathrm{mT}$ antiparallel to the $x$-axis (middle), and $15~\mathrm{mT}$ parallel to the $\hat{u}_3$ unit vector, that is rotated by $22.5^\circ$ clockwise with respect to the $y$-axis.}
	\label{figS10}
\end{figure}

Next, the in-plane field was applied along the $x$-direction, as displayed in Figs.~\ref{figS10}(a),(b),(d),(e),(g),(h). Unlike the case of $B_y$, the asymmetry due to $B_x$ was very small, although still present both in the constant-bias cut planes and in the spectra, and flipped upon field inversion. For $B_x >0$, the asymmetry was in the same direction as for $B_y < 0$: this suggests that the direction orthogonal to $\BSO$, where the asymmetry is expected to be completely absent, was close to the $x$-direction and, for a certain $B_x > 0$ component, had a smaller $B_y > 0$ component. Following this argument, the vector $\hat{u}_2$ was defined $22.5^\circ$ rotated counterclockwise with respect to $B_x$, targeting a direction nearly orthogonal to $\BSO$. As shown in Figs~5(c) and 5(f) of the Main Text, the asymmetry along $\hat{u}_2$ was very small (smaller than for $B_x$). Consequently, the vector $\hat{u}_1$ was defined orthogonal to $\hat{u}_2$ to be approximately aligned with $\BSO$, indeed revealing the largest asymmetry for given $|\mathbf{B}|$ [Figs.~5(a),(b),(d),(e)].

Finally, we investigated the effect of the magnetic field applied along an intermediate direction between $\hat{u}_1$ and $\hat{u}_2$, defined as $\hat{u}_3 \equiv (\hat{u}_1 + \hat{u}_2)/\sqrt{2}$ (i.e., rotated $22.5^\circ$ degrees clockwise with respect to $B_y$). The result is shown in Figs.~\ref{figS10}(c),(f),(i): the asymmetry was visible in both the constant-bias planes and in the spectrum, although smaller than for the same $|\mathbf{B}|$ applied along $\hat{u}_1$, compatible with the smaller field component along $\BSO$. As expected, the direction of the asymmetry was the same as for $\mathbf{B} \cdot \hat{u}_1 > 0$.

\setcounter{myc}{10}

\section{Results for Device 2}

\setcounter{myc}{11}
\begin{figure}[h!]
	\includegraphics[width=0.7692\columnwidth]{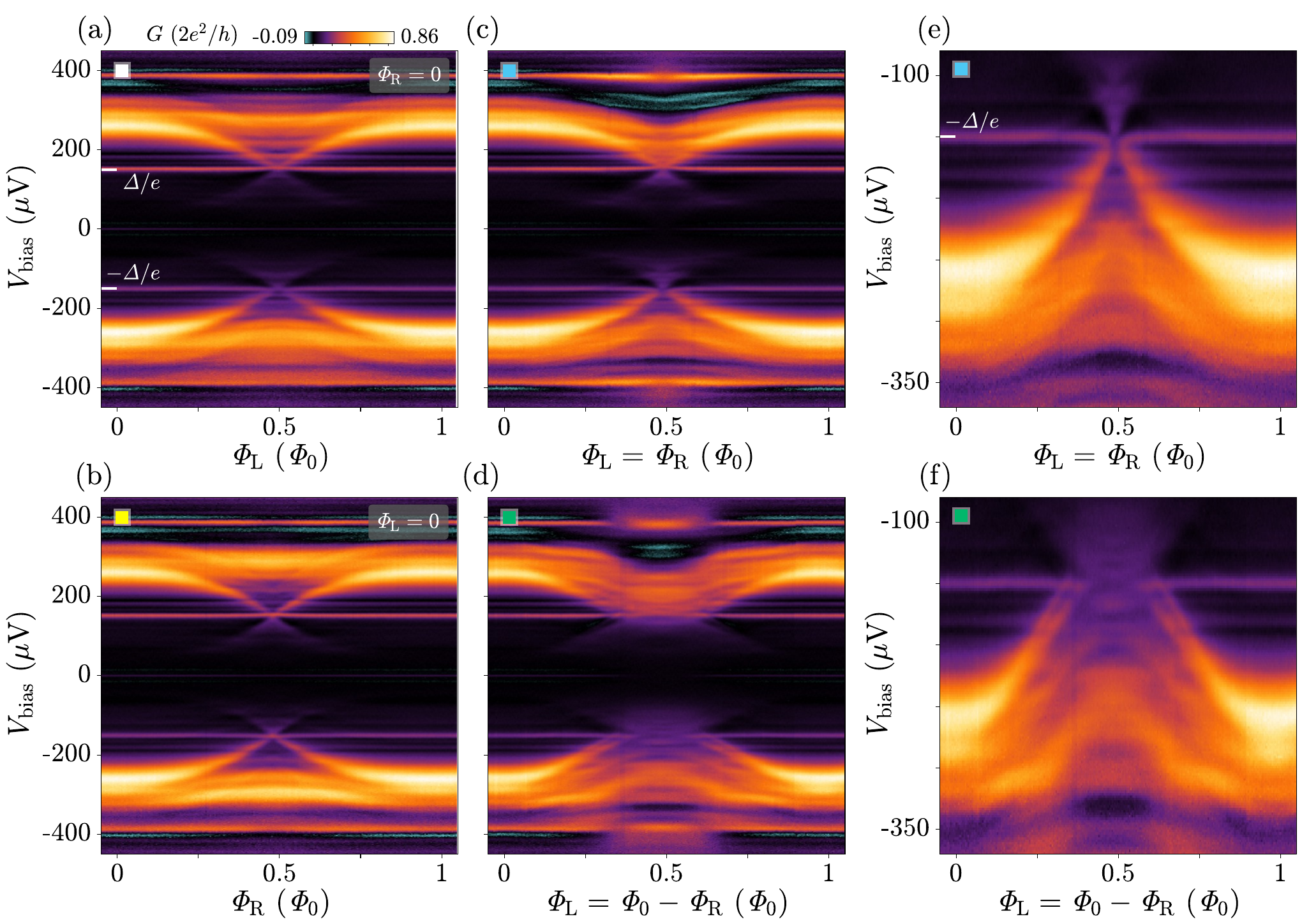}
	\caption{\textbf{Tunneling spectroscopy along phase-space linecuts for Device 2.}
		(a) Differential conductance $G$ as a function of voltage bias $\Vbias$ and flux $\PhiL$, for $\PhiR=0$ [white line in Fig.~\ref{figS12}(a)].
		(b) As in (a), but as a function of $\PhiR$, for $\PhiL=0$ [yellow line in Fig.~\ref{figS12}(a)].
		(c) As in (a), but along the phase-space linecut $\PhiL = \PhiR$ [dashed cyan line in Fig.~\ref{figS12}(a)].
		(d) As in (a), but along the phase-space linecut $\PhiL = \Phio - \PhiR$ [dashed green line in Fig.~\ref{figS12}(a)].
		(e),(f) Zoom-in of (c),(d).}
	\label{figS11}
\end{figure}

\setcounter{myc}{12}
\begin{figure}[h!]
	\includegraphics[width=0.5\columnwidth]{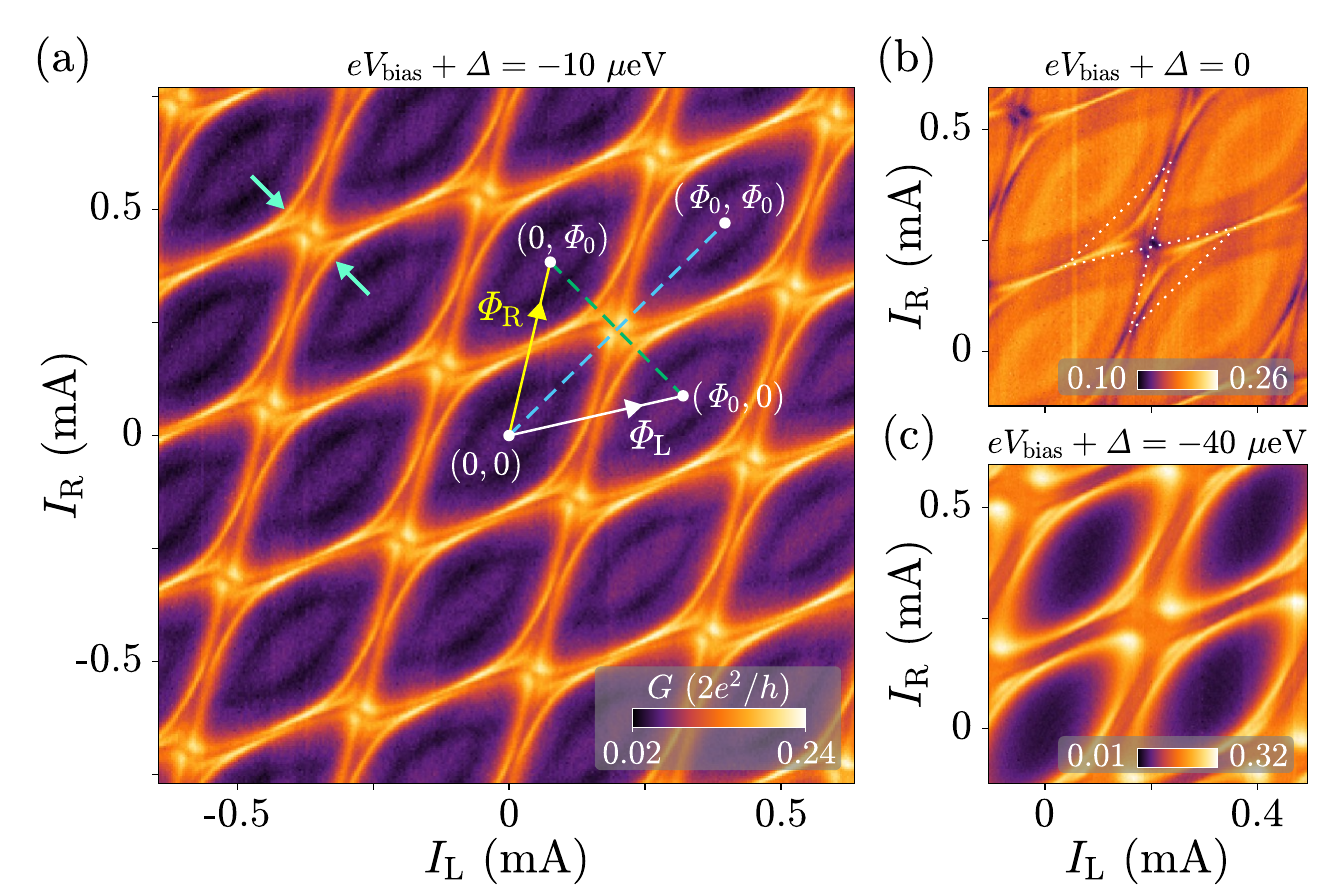}
	\caption{\textbf{Constant-energy cut planes in the two-dimensional phase space for Device 2.}
		(a) Differential conductance $G$ as a function of flux-line currents $\Il$ and $\Ir$, for $\Vbias = -161~\mathrm{\mu V} = -\Deltait/e - 10~\mathrm{\mu V}$. Solid white and yellow lines indicate the periodicity directions, corresponding to the external magnetic fluxes $\PhiL$ and $\PhiR$. Dashed cyan and green lines represent the $\PhiL = \PhiR$ and $\PhiL = \Phio - \PhiR$ linecuts. Turquoise arrows indicate the split resonances discussed in the text.
		(b) As in (a), but for $\Vbias = -151~\mathrm{\mu V} = -\Deltait/e$. Dotted white triangles show the perimeter of regions where the discrete vortex condition is fulfilled.
		(c) As in (a), but for $\Vbias = -191~\mathrm{\mu V} = -\Deltait/e - 40~\mathrm{\mu V}$.}
	\label{figS12}
\end{figure}

Tunneling spectroscopy measurements were performed on a second device, fabricated on the same chip of the first and measured in the same cool down. The only difference between Device 1 and 2 was the size of the superconducting island in the middle of the scattering region, whose diameter was enlarged from $200~\mathrm{nm}$ to $300~\mathrm{nm}$ for Device 2.
In Figs.~\ref{figS11} and \ref{figS12}, we report the main results obtained for Device 2 at zero magnetic field. Here, the tunnel gates and the probe gate were set to $\Vtl=\Vtr=-1.38~\mathrm{V}$ and $\Vp = 0.3~\mathrm{V}$ to operate in regime comparable to that shown for Device 1. The left, middle and right gates were set to $\Vl = \Vm = \Vr = 0$. Bias-dependent spectroscopy, displayed in Fig.~\ref{figS11}, featured a transport gap related to the superconducting probe of $2 \Deltait / e = 302~\mathrm{\mu V}$, slightly lower than for Device 1 but still compatible with the superconducting gap of Al, and distinct flux-independent conductance peaks at $\pm \Deltait /e $ due to MAR. 
The spectrum is first shown as a function of $\PhiL$ for $\PhiR = 0$ [panel (a), corresponding to Fig.~\ref{figS2}(a) for Device 1], as a function of $\PhiR$ for $\PhiL = 0$ [panel (b), corresponding to Fig.~\ref{figS2}(b)] and along the $\PhiL = \PhiR$ linecut [panels (c) and (e), corresponding to Figs.~3(a) and (c) of the Main Text]. In these three cases, the flux-dependent ABSs resembled the conventional energy--phase dispersion valid for two-terminal JJs \cite{Beenakker1991b}, as one phase difference is zero. Notably, ABSs approaching the probe gap edges indicate near-unity transmission of the respective conduction channels. Spectroscopy performed along the $\PhiL = \Phio - \PhiR$ linecut [Figs.~\ref{figS11}(d) and \ref{figS11}(e)] revealed the two main features highlighted for Device 1, namely splitting of Andreev levels and conductance enhancement at the MAR peaks marking zero-energy Andreev level crossings, that were interpreted as lifting of the ABS spin degeneracy and ground state fermion parity transitions. The size of the splitting was smaller than in Device 1 by approximately a factor 2. This may be attributed to the larger size of the scattering region (here $400~\mathrm{nm} \times 350~\mathrm{nm}$) on the scale of the superconducting coherence length in InAs $\xi_\mathrm{InAs} \sim 600~\mathrm{nm}$, and to device-to-device variability.

Selected constant-bias cut planes for Device 2 are presented in Fig.~\ref{figS12} (to compare with Fig.~4 of the Main Text). The scan over an extended region of the phase space [Fig.~\ref{figS12}(a), taken at $e \Vbias = - \Deltait - 10~\mathrm{\mu e V} = -161~\mathrm{\mu e V}$] showed periodicity along the $\PhiL$- and $\PhiR$-axes. The linecuts relevant for Fig.~\ref{figS11} are plotted (color coded). Split lines along $\PhiL = \Phio - \PhiR$, identified by the turquoise arrows, correspond to the spin-split levels observed in Figs.~\ref{figS11}(d) and \ref{figS11}(f). By reducing $|\Vbias|$ to probe the phase space at zero energy [Fig.~\ref{figS12}(b)], we confirm the presence of odd-parity triangular regions whose perimeter is marked by conductance resonances (that is, zero-energy crossings). Such regions are within the portions of phase space where the discrete vortex condition is fulfilled, shown as the dotted white triangles. Finally, the constant-bias cut plane at $\Vbias = -191~\mathrm{\mu V}$ ($40~\mathrm{\mu eV}$ below zero energy) is displayed in Fig.~\ref{figS12}(c), showing additional ABSs and the split lines at a limit of visibility.
The reproducibility of our experimental observations on a second device suggests the generality of these phenomena and further support our interpretation.

\end{document}